\documentclass[twocolumn,prl,longbibliography,superscriptaddress]{revtex4-2}
 
\usepackage{amsthm,amsmath,amssymb,amsfonts,graphicx,verbatim,xcolor,bm,physics} 
\usepackage{ulem} 
\usepackage[utf8]{inputenc} 
\usepackage{siunitx} 
\usepackage[colorlinks=true]{hyperref}

\newcommand{\beqn}{\begin{eqnarray}}
\newcommand{\eeqn}{\end{eqnarray}}

\newcommand{\br}{\bm r}
\newcommand{\bk}{\bm k}
\newcommand{\bq}{\bm q}

\renewcommand{\v}[1]{\ensuremath{\mathbf{#1}}} 

\begin{document}

\title{Layer skyrmions for ideal Chern bands and twisted bilayer graphene}


\author{Daniele Guerci} 
\affiliation{Center for Computational Quantum Physics, Flatiron Institute, New York, New York 10010, USA}
\author{Jie Wang} 
\affiliation{Department of Physics, Temple University, Philadelphia, Pennsylvania, 19122, USA}
\author{Christophe Mora} 
\affiliation{Universit\'e Paris Cit\'e, CNRS,  Laboratoire  Mat\'eriaux  et  Ph\'enom\`enes  Quantiques, 75013  Paris,  France}

\begin{abstract}
Ideal $C=1$ Chern bands exhibit a Landau level correspondence: they factorize as a lowest Landau levels and a spinor wavefunction that spans the layer index. We demonstrate that, in single Dirac moiré models, the spinor develops generally a Skyrme texture in real space with an associated Berry phase which compensates exactly the magnetic phase of the Landau level. For ideal bands with higher Chern numbers $C>1$, we find that $C$ color Landau levels are carried by $C$ spinors with Skyrme textures. We identify a SU(C) gauge symmetry in the color space of spinors and an emergent non-Abelian connection in real space intimately linked to the Pontryagin winding index of the layer skyrmions. They result in a total real-space Chern number of $-1$, screening the magnetic phase, irrespective of $C$ and of the number of layers. The topologically robust Skyrme texture remains remarkably intact in twisted bilayer graphene, even far from the chiral limit, and for realistic values of corrugation, making it an experimentally testable feature. We verify our predictions at the first magic angle of twisted bilayer, trilayer, and monolayer-bilayer graphene.


\end{abstract}

\maketitle

\paragraph*{Introduction ---} 
The connection between Chern bands and Landau levels~\cite{Neupertprl_2011,PhysRevLett.106.236802,PhysRevX.1.021014,Liu_2012,Yang2012,Wu_2013,PhysRevLett.105.215303,ModelFCI_Zhao,zhao_nonAbelian,zhao_review,Liu_2024} has recently sparked renewed interest~\cite{jain2023twist,repellin2023twisted,NicolasMD_2024}, stimulated by the experimental observation of integer and fractional quantum anomalous Hall phases~\cite{Sharpe605,Serlin900,FCI_TBG_exp,Cai2023,PhysRevX.13.031037,makKangEvidenceFractionalQuantum2024,xuParkObservationFractionallyQuantized2023,lu2023fractionalquantumanomaloushall,zeng2023thermodynamic,xie2024evenodddenominatorfractionalquantum} in moir\'e materials. 
Unlike Landau levels, Chern bands are characterized by a non-uniform Berry curvature and finite dispersion, making their comparison with Landau levels challenging~\cite{Goerbig_2012,PARAMESWARAN2013816,ParameswaranPRB_2012,Roy_PRB_2014,liu2024theorygeneralizedlandaulevels}. 
Moreover, since Chern bands do not originate from an external real-space magnetic field, fundamental questions arise about establishing a direct correspondence with Landau levels and implementing flux attachment techniques~\cite{hlr_1993,murthy2011composite,Barkeshli_2012,ledwith2022vortexability,Estienne_2023}. Remarkably, recent advancements in understanding flat bands in graphene-based moir\'e materials have introduced a class of ideal wavefunctions, and show them to be exactly expressed as the product of a Lowest Landau level (LLL) and a layer vector describing the electronic distribution across different layers~\cite{Grisha_TBG,kahlerband1,kahlerband2,kahlerband3,PhysRevLett.124.106803,PhysRevB.108.075126,sarkar2023symmetrybased,KaiSunPRL_2023,Le_2024,Grisha_TBG2,Wang_2021,JieWang_exactlldescription}.
\begin{figure}
    \centering
    \includegraphics[width=0.8\linewidth]{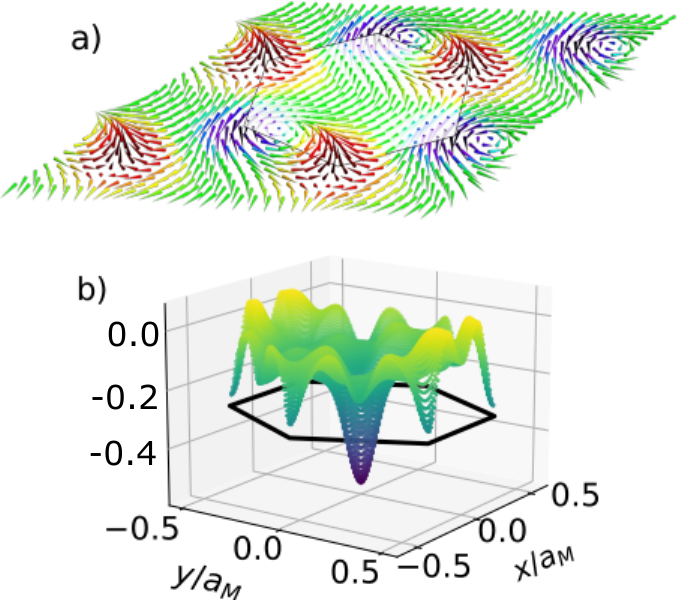}
    \caption{a) Layer skyrmion of chiral TBG  at the first magic angle, with the moiré unit cell indicated in black.  b) Pontryagin density $Q(\v r)$ (integrand of Eq.~\eqref{winding-pontryagin}) for the CP$^2$ layer skyrmion of the $C=1$ ideal band of helical trilayer graphene. }    
    \label{fig:skyrmions_C1}
\end{figure}

The Landau level correspondence of Chern bands has also been addressed from a different perspective within the adiabatic approach to the continuum model of twisted transition metal dichalcogenides (TMD)~\cite{DongPRL_2023,GoldmanPRL_2023,PhysRevLett.132.036501,NicolasMDPRL_2024,shi2024adiabatic,li2024contrasting,reddy2024nonAbelian}, such as twisted homobilayer MoTe$_2$  ($t$MoTe$_2$)~\cite{FWu_PRL_2019,NicolasMDPRL_2024,shi2024adiabatic,zhang2024polarizationdriven,wang2024higher}. 
There, the wavefunction factorizes into a layer skyrmion and a lattice model with an effective magnetic field. This inhomogeneous magnetic field originates from the Berry phase of the skyrmion winding in real space~\cite{NicolasMDPRL_2024,shi2024adiabatic}. Although the wavefunction is in general not an ideal one, it seemingly becomes ideal when the band flattens~\cite{DongPRL_2023,Cr_pel_2024,shi2024adiabatic}.


In this letter, we unite these two perspectives and demonstrate quite generally that the spinor of ideal Chern bands always develops a Skyrme texture in single-Dirac moiré models.
The skyrmion winding, characterized by the topologically quantized Pontryagin index, generates a space-dependent Berry phase necessary to cancel the fictitious magnetic phase of the LLL and form a Chern band. For twisted bilayer graphene (TBG), we illustrate our findings for the ideal $C=1$ Chern band in the chiral limit at the first magic angle. Remarkably, the Skyrme texture is not at all restricted to the academic chiral limit; it persists up to realistic values of corrugation in TBG due to its topological robustness. In the case of ideal bands with higher $C$ Chern numbers~\cite{wang2023origin,dong2022,mera2023uniqueness,Estienne_2023}, we extend our decomposition by introducing $C$ color LLLs and their associated color skyrmions. They transform under SU(C) with a non-abelian gauge freedom related to the choice of LLL magnetic unit cell. The gauge-invariant topological winding of the skyrmions in real space again screens the LLL magnetic phase.


\paragraph*{Layer skyrmions in $C=1$ ideal bands ---} The canonical model hosting $C=1$ ideal flatbands is the chiral limit of TBG~\cite{Grisha_TBG,Grisha_TBG2}.
The wavefunction in the chiral sector is expressed in terms of a generalized LLL $\Phi_{\bk}(\br)$ experiencing an inhomogeneous magnetic field of one flux quantum 
per moir\'e unit cell, and a layer spinor $\bm\chi(\br)$ quantifying the electronic distribution across the two layers: 
\begin{equation}\label{ideal_C1_wfc}
    \bm\psi_{\bk}(\br)=\bm\chi(\br)\,\Phi_{\bk}(\br).
\end{equation}
The LLL wavefunction carries the Chern $C=1$ character of the band corresponding to a Berry  flux in momentum space, dual to a real-space fictitious magnetic field of magnetic length $\ell_B$. The Bloch translation symmetry of the band wavefunction does not match the magnetic phase of the LLL, $\Phi_{\bk}(\br+\bm a_j)=e^{i(\bm a_j\times \br)/2l^2_B}e^{i\bk\cdot\bm a_j}\Phi_{\bk}(\br)$, with the unit cell vectors $\v a_1$ and $\v a_2$. This discrepancy must be compensated by a screening phase from the spinor wavefunction $\bm\chi(\br)$, similar to an antiquantum Hall state~\cite{Wang_2021}. 
For a single-component function $\chi(\br)$, realizing the screening boundary condition (SBC) $\bm \chi (\br+\bm a_j)=e^{-i(\bm a_j\times \br)/2l^2_B}\bm \chi (\br)$ is equivalent to having a quantized vortex-like winding of its phase by $-2 \pi$ around the moiré unit cell. As $\chi(\br)$ is single-valued, this is only possible if $\chi(\br)$ has at least one zero. As $\bm\chi(\br)$ does not vanish for single-Dirac moiré models~\footnote{Models with quadratic band touching~\cite{Eugenio_2023,Wan_prl_2023} provide an example where $\chi$ vanishes, allowing for single-layer screening~\cite{Suppmat}.}, and we set $|\bm \chi|=1$, this proves that the magnetic screening requires the spinor $\bm\chi(\br)$ to possess at least two components and, as a result, ideal bands cannot form in single-layer models of Dirac electrons~\cite{gaogiang2023,Wan_prb_2023}. 

Next, for two layers, we prove that the Pontryagin index of the gauge invariant CP$^1$ layer skyrmion $\bm n=\bm\chi^\dagger\bm\sigma\bm\chi$, illustrated in Fig.~\ref{fig:skyrmions_C1}a) for TBG, and the real space Chern number $C_R$ coincide. The Pontryagin index
\begin{equation}\label{winding-pontryagin}
    \mathcal W_{\chi} = \frac{1}{4 \pi}\int_{\rm UC}  d^2\br \, \, \bm n\cdot(\partial_x\bm n\times\partial_y\bm n) 
\end{equation}
indicates the covering of the unit sphere by $\bm n (\br)$
over a moiré unit cell. The skyrmion winding texture generates a Berry connection 
$\bm {\mathcal A}=-i\bm \chi^\dagger\nabla_{\br}\bm\chi$ in real space, associated with $C_R =  \mathcal W_{\chi}$ as detailed in the Supplemental Material (SM)~\cite{Suppmat}. We have verified the equality for the first magic angle of chiral TBG, whose skyrmion texture is represented Fig.~\ref{fig:skyrmions_C1}a), by evaluating
\begin{equation}\label{screening}
    C_R =\mathcal W_{\chi}= -1 
\end{equation}
Eq.~\eqref{screening} extends to larger number of layers $L>2$.
The gauge invariant layer skyrmion $n_a=\bm \chi^\dagger\lambda_a\bm\chi$ belongs to CP$^{L-1}$ with the  generators $\lambda_a$ of SU(L). The topological winding follows Eq.~\eqref{winding-pontryagin} by extending the triple product with the SU(L) structure factors~\cite{PhysRevResearch.4.023120,PhysRevB.104.085114}. It is also equal to the real space Chern number $C_R$ calculated from the Wilson loop accumulated by the spinor $\bm\chi(\br)$.
As a representative example, we confirm numerically Eq.~\eqref{screening} for a chiral model of helical twisted trilayer graphene (hTTG) hosting a $C=1$ ideal flat band with three layers~\cite{guerci2023chern,guerci2023nature}. 
Interestingly, the skyrmion texture is reminiscent of the adiabatic approach used to describe bands in twisted TMDs~\cite{DongPRL_2023,FWu_PRL_2019,NicolasMDPRL_2024,shi2024adiabatic,zhang2024polarizationdriven,wang2024higher}, although such bands are generally not ideal. Remarkably, the Berry phase of the skyrmion texture exactly compensates  the magnetic phase of the LLL and recovers Bloch periodicity.

\begin{figure}
    \centering
    \includegraphics[width=\linewidth]{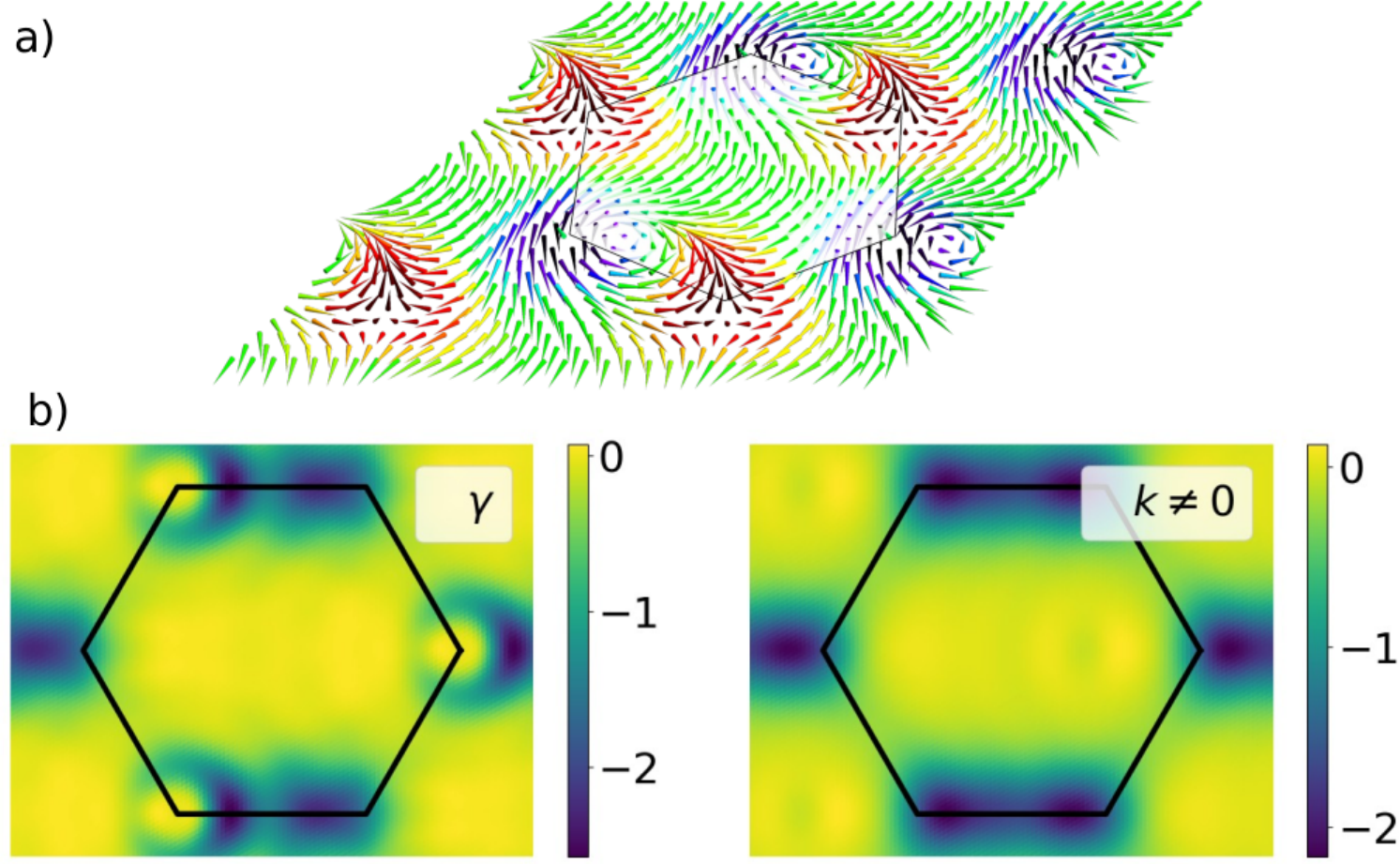}
    \caption{Layer skyrmions of TBG for realistic atomic corrugation with $w_{\rm AA}/w_{\rm AB}=0.7$. a) Vector field $\bm n_{\bm k}$ at the $\gamma$ point. b) Pontryagin densities at $\gamma$ and $\bk = -(\v b_1+ \v b_2)/4$ over the moiré lattice, both integrating to $W_{\chi,\bk}=1$. Deviations from the chiral limit are enhanced at $\gamma$ due to a stronger mixing to the remote bands~\cite{Bistritzer_2011,Zhida_PRL19,Oscar_hiddensym,hfm_song_2022}. Black lines show the moir\'e unit cell.}   
    \label{fig:p_density_away}
\end{figure}

\paragraph*{Layer skyrmions for realistic TBG ---}  
The Skyrme texture is topologically robust and persists away from the chiral limit. There, 
the two Chern bands couple and the factorization Eq.~\eqref{ideal_C1_wfc} no longer holds but it is still possible to diagonalize the sublattice operator within the two central bands~\cite{Zaletel_PRX20} at each $\bk$, and thus continuously define $\bm\psi_{\bk}(\br)$ as the positive sublattice-polarized state. The resulting band $\bm\psi_{\bk}(\br)$ does not diagonalize kinetic energy but it has a well-defined $C=1$ Chern number. In a strong coupling approach~\cite{Oskar_PRL19,Zaletel_PRX20,Ledwith_ann_2021,Oscar_hiddensym,lian2020tbg}
where interactions favor a valley-spin-Chern polarized state, electrons populate maximally the band formed by $\bm\psi_{\bk}(\br)$ and realize a quantum anomalous Hall effect at integer fillings~\cite{Serlin900,nuckolls2020strongly,Stepanov_2021,Sharpe605}. We define the normalized spinor $\bm\chi_{\bk}(\br) = \bm\psi_{\bk}(\br)/|\bm\psi_{\bk}(\br)|$ and the related CP$^1$ vector 
 $\bm n_{\bk}=\bm\chi_{\bk}^\dagger\bm\sigma\bm\chi_{\bk}$
which both acquire a momentum $\bk$ dependence.

Fig.~\ref{fig:p_density_away} illustrates the results for the Skyrme texture and the Pontryagin densities for the realistic ratio $w_{\rm AA}/w_{\rm AB}=0.7$~\cite{Zhida_PRL19} of tunnelings in TBG. Despite the $\bk$-dependence and the real space strong variations, the Pontryagin index $\mathcal W_{\chi,\bk}$ remains quantized to $-1$ for all $\bk$ which evidences the robustness of the layer spinor winding away from the chiral limit. This stable feature of ideal flat bands may serve as as starting point when decomposing non-ideal topological bands into higher Landau levels~\cite{liu2024theorygeneralizedlandaulevels,Fujimoto_higher_vortexability24,li2024variational}. We note that, by time-reversal symmetry, the negative sublattice-polarized $C=-1$ band develops an opposite Skyrme texture with $\mathcal W_{\chi,\bk} = +1$. For homobilayer twisted TMDs, non-ideal Chern bands are expected to exhibit Skyrme textures even beyond the adiabatic approximation~\cite{NicolasMDPRL_2024}, as they are continuously connected to an ideal limit~\cite{DongPRL_2023}. 



\paragraph*{SU(C) structure of periodic Landau levels---}
Before turning to the skyrmion texture of higher Chern bands, we take a detour and discuss an emergent SU(C) symmetry for lattice periodic LLL~\cite{Wu_2013,Barkeshli_PRX_2012,wu2014,dong2022,wang2023origin}.
We consider a lattice $(\v a_1, \v a_2)$ commensurate to an external homogeneous magnetic field where a single flux quantum pierces $C$ unit cells. We arbitrarily choose 
$(C\v a_1,\v a_2)$ for the magnetic unit cell and the  magnetic translations across the lattice satisfy $T_{\v a_1}T_{\v a_2}=e^{i\phi}T_{\v a_2}T_{\v a_1}$ with $\phi=2\pi/C$, along with the commutation of  $T_{C\v a_1}$ and $T_{\v a_2}$. 
With $\v R$ the guiding center position, the LLL states  $\ket{\bk}=e^{i\bk\cdot\v R}\ket{0}$ satisfy Bloch periodicity over the extended magnetic lattice by diagonalizing both magnetic translation operators $T_{C\v a_1}$ and $T_{\v a_2}$.
For each momentum $\bm k$, we shall define a set of $C$ color LLL wavefunctions and examine the representation of the magnetic translation in this color basis.
We define the first color component as: 
\begin{equation}
    F^{z}_{\bk,1}(\br) = \braket{\br}{\bk}= U^{z}_{\bk,1}(\br) e^{i\bk\cdot\br},
\end{equation}
which satisfies $T_{C\v a_1}U^{z}_{\bk,1}(\br)=U^{z}_{\bk,1}(\br)$ and $T_{\v a_2}U^{z}_{\bk,1}(\br)=U^{z}_{\bk,1}(\br)$. The superscript $z$ refer to our specific choice of diagonalizing  $T_{\v a_2}$. 
The subsequent color LLL are obtained by applying the magnetic translation operator: 
\begin{equation}
    U^{z}_{\bk,n}(\br)=T_{\bm a_1} U^{z}_{\bk,n-1}(\br)
\end{equation}
corresponding to $C$ different colors due to the cyclic condition that $U^{z}_{\bk,1}$ is invariant under $T_{C\bm a_1} = (T_{\bm a_1})^C$. We regroup all colors in a single vector $\bm U^{z}_k=[U^{z}_{k,1},\dots,U^{z}_{k,n}]$ to present the magnetic translations $T_{\bm a_1}$, $T_{\bm a_2}$ respectively as the matrices
\begin{equation}
    \sigma = \begin{pmatrix}
        0 & 1 & \ldots  & 0 \\
        \vdots  & \ddots &  \ddots & \vdots \\
        0 &   \ldots & \ldots & 1  \\
        1 &  0 & \ldots &  0
    \end{pmatrix},\, \tau =\text{diag}[1,e^{i\phi},\cdots,e^{i(C-1)\phi}], 
\end{equation} corresponding to the algebra of  clock or parafermion models~\cite{BAXTER1989155,Fendley_2012,Fendley_2014,Yangle_Modelwf},
with $\sigma^C = \tau^C = 1$ and the commutation relation $\sigma \tau = e^{2 i \pi/C} \tau \sigma $. The third magnetic translation $T_{\bm a_3}$ along $\bm a_3=-\bm a_1-\bm a_2$ is such that  $T_{\bm a_1}T_{\bm a_2}T_{\bm a_3}=e^{i\pi/C}$, imposed by the flux threading the triangle formed by $\bm a_{1,2,3}$. 
Combining the different powers of the magnetic translation operators $T_{\bm a_1}$ and $T_{\bm a_2}$ generates $C^2-1$ traceless generators of SU(C).

The matrices $\sigma$ and $\tau$ play a symmetric role and share the same set of eigenvalues. Their specific form is a gauge fixing and depends on the direction where we extend the unit cell. For instance, 
diagonalizing $T_{\bm a_1}$ implements a rotation from $\bm U^z_{\bk}$ to $\bm U^x_{\bk}$ for which the role of $\sigma$ and $\tau$ are interchanged:
\begin{equation}
   T_{\bm a_1} \bm U^{x}_{\bk}=\tau \bm U^{x}_{\bk},\, \quad T_{\bm a_2} \bm U^{x}_{\bk}=\sigma \bm U^{x}_{\bk}.
\end{equation}
In particular, a diagonal $T_{\bm a_1}$ indicates a magnetic unit cell $(\bm a_1,C\bm a_2)$, rotated with respect to the original one.
Similarly, $\bm U^y_{\bk}$ diagonalizes the boundary condition along $\bm a_3$ leading to a magnetic unit cell $(\bm a_1,C\bm a_3)$.
Table~\ref{tab:color_basis} lists different representations of the magnetic translation depending on the choice of color basis for $C=2$.
Remarkably, rotating through the color basis redefines the magnetic unit cell without changing its area. 

By construction, each periodic color $F^{z}_{\bk,1}$ has Chern number $1$ over the magnetic unit cell $(\v b_1/C,\v b_2)$. The Berry flux being additive, it is augmented to Chern number $C$ in the reciprocal unit cell spanned by $\v b_1$ and $\v b_2$. It already indicates that these states provide a suitable basis for Chern $C$ ideal bands. Finally, the transport across the unit cell combines a magnetic phase and a rotation in color space,
$\bm F_k^z (\br+\bm a_{1/2}) = e^{i(\bm a_j\times \br)/2l^2_B}  (\sigma/\tau) \bm F_k^z (\br)$.
\begin{table}[]
\centering
\begin{tabular}{|c||c|c|c|}
\hline
 Basis & $T_{\bm a_1}$ & $T_{\bm a_2}$ & $T_{\bm a_3}$    \\
\hline\hline
$\bm U^z_k(\br)$ &  $\mu^x$  & $\mu^z$  & $-\mu^y$  \\
\hline
$\bm U^x_k(\br)$ & $\mu^z$  &  $\mu^x$ & $\mu^y$ \\
\hline
$\bm U^y_k(\br)$ & $-\mu^y$  & $\mu^x$ & $\mu^z$   \\
\hline
\end{tabular}
\caption{Representation of magnetic translations for different choices of color basis with $C=2$.
We show three representative directions, $\v a_1$, $\v a_2$ and $\v a_3$ and use the Pauli matrices $\mu^a$ in color space.}
\label{tab:color_basis}
\end{table}

\paragraph*{Layer skyrmions in higher Chern bands---} 
We now show that ideal Chern $C$ bands generally decompose as
\begin{equation}\label{color_decomposition}
    \bm \psi_{\bk}(\br)= \sum_{j=1}^C \bm\chi_j^z(\br)\,  F^z_{\bk,j}(\br) e^{-K_j(\v r)} ,
\end{equation}
a sum over $C$ spinors $\bm\chi_j^z$ weighed by the color LLL wavefunctions $F^z_{\bk,j}$. Eq.~\eqref{color_decomposition} fits into the general form of the wavefunction in Ref.~\cite{wang2023origin} determined from holomorphicity and translation symmetry.
The real (Kähler) functions $K_j (\br)$  are introduced to normalize the spinors and identify as color-dependent inhomogenities in the fictitious magnetic fields that are periodic and average to zero over the moiré lattice.
The scalar product form of Eq.~\eqref{color_decomposition} implies a gauge freedom, allowing both color and layer spinor wavefunctions to be rotated simultaneously, from $z$ to $x$ for instance. The gauge structure induces an  
SU(C) non-Abelian gauge field 
\begin{equation}\label{berry-connection}
\bm {\mathcal A}_{nm}=-i\tilde{\bm \chi}^{z \dagger}_n\nabla_{\br}\bm \chi^z_m    
\end{equation}
which generalizes the real-space Berry connection to $C>1$ bands. It involves the dual basis $\tilde{\bm \chi}^{z \dagger}_n=X^{-1}_{nm}{\bm \chi}^{z \dagger}_m$ with $X_{nm}={\bm \chi}^{z \dagger}_{n}\cdot{\bm \chi}^z_m$ accounting for the non-orthogonality of the different spinor wavefunctions. Similarly to the $C=1$ case, the magnetic translation properties of the LLL must be screened by the non-diagonal boundary conditions
\begin{equation}\label{real-space-BC}
    \bm\chi^z(\br +\v a_{1/2}) = e^{i(\bm r\times \v a_j)/2l^2_B}   \bm\chi^z(\br) (\sigma^\dagger/\tau^\dagger)
\end{equation}
on the spinors. With the gauge choice along $z$, the boundary conditions are again diagonal over the extended magnetic unit cell $(C \v a_1,\v a_2)$, but keep a magnetic phase. Each spinor thus develops a skyrmion texture with a Pontryagin winding index of $-1$ over the magnetic unit cell, also obtained by integrating the Berry connection Eq.~\eqref{berry-connection} along the edges of this extended cell. Due to Eq.~\eqref{real-space-BC}, the gauge field can also be decomposed as 
\begin{equation}
    \bm {\mathcal A} (\v r) = \frac{\br\times\bm z}{2l^2_B}\lambda_0 + \bm {\mathcal A}_{0}(\br)\lambda_0
    +\sum^{C^2-1}_{j=1} \bm {\mathcal A}_{j}(\br)\lambda_j,
\end{equation}
with $\bm {\mathcal A}_{0} (\v r)$ periodic over $\v a_j$, $\lambda_j$ are the generators of SU(C) and $\lambda_0$ the identity.
The first two terms represent the Abelian part while the third one originates from the SU(C) non-Abelian color structure. Taking the trace removes all non-Abelian components and the real space Chern number $C_R$, obtained by integrating around the original unit cell $(\v a_1,\v a_2)$, equals $-1$ with a contribution of $-1/C$ given by each color skyrmion. It completes the screening of the $C$ color LLL wavefunctions, each carrying a flux of $1/C$ over the unit cell. One interesting outcome of this structure of ideal Chern bands is that, as shown in the SM~\cite{Suppmat}, in order to develop a non-trivial Wilson loop topology, the $C$ spinors must evolve in a layer space larger than their number.  
It directly implies that $C \le L-1$, extending the result that ideal Chern $C=1$ bands require at least two layers of Dirac electrons to develop.

 \begin{figure}
    \centering
    \includegraphics[width=0.9\linewidth]{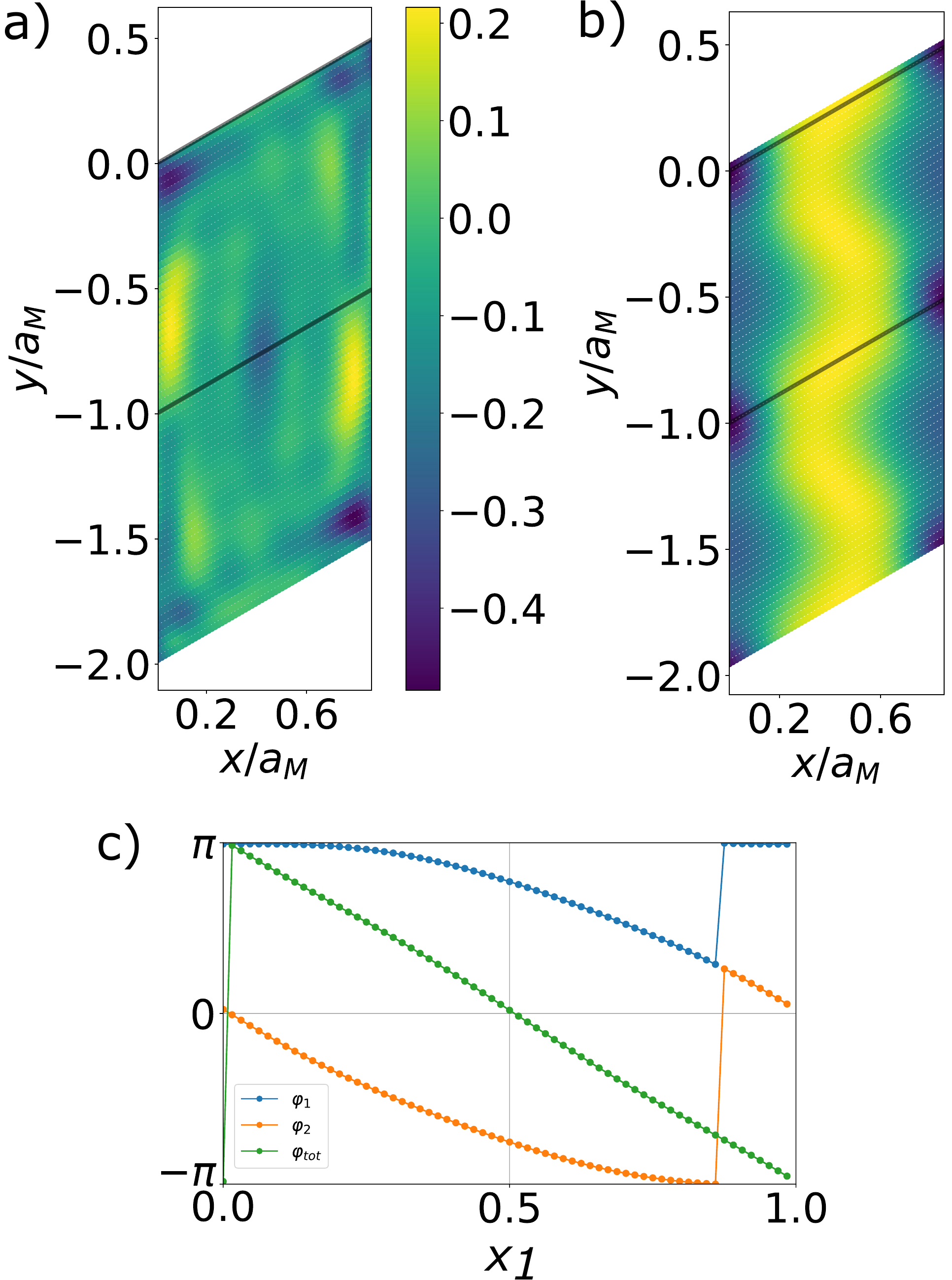}
    \caption{a) Pontryagin density $Q_1 (\bm r)$ for the first color skyrmion with winding $-1$ per magnetic unit cell $(2 \v a_1,\v a_2)$. b) 
    Abelian part of the Berry curvature $\Omega=\Tr[\nabla_{\bm r}\times\bm{\mathcal A}]$.
    c)  Wilson loop eigenvalues along $\v a_2$ with $x_1=\v b_1\cdot\br/2\pi$. A moiré cycle swaps the two eigenvalues while their sum winds by $2 \pi$.   
    Calculations are done for the $C=2$ ideal flat band of hTTG at the magic angle $\theta= 1.687^\circ$.}    
    \label{fig:figure2}
\end{figure}

\paragraph*{$C=2$ ideal bands in helical trilayer graphene ---} We illustrate the above ideas with a specific example. The chiral model of helical twisted trilayer graphene exhibits an ideal Chern band with $C=2$ in an ABA stacking configuration at the first magic angle. The zero mode wavefunction has the analytical expression~\cite{guerci2023chern,guerci2023nature}
\begin{equation}\label{flatAband}
   {\bm \psi}_{\bk}(\br) =  {\cal N}_k e^{\frac{\pi}{\rm Im \omega} (k^2 - |k|^2)} \sum_{\pm} 
   a_{\pm k} 
   f_{\bk \mp \v K} (z) {\bm \psi}_{\pm K}(\br).
\end{equation}
with a normalization factor ${\cal N}_k$.
 $\psi_K (\br)$ and $\psi_{-K}(\br)$ are symmetry-protected zero modes at arbitrary twist angle whereas the function~\cite{haldanetorus1,haldanemodularinv,Grisha_TBG,Grisha_TBG2}
\begin{equation}\label{holomorphic}
   f_{\bk} (z) = e^{i \v k \cdot \v a_1 z} \vartheta_1[ z - k,\omega]/\vartheta_1[ z,\omega] 
\end{equation}
    is holomorphic in $z$, with $\omega= a_2/a_1$, the Weierstrass pseudo-periodic $\vartheta_1$ function and $a_k  = \vartheta_1 (k+K,\omega)$. $z$ and $k$ are the complex numbers associated to $\v r$ and $\v k$ and divided respectively by $a_1$ and $b_2$. As shown in Refs.~\cite{guerci2023chern,guerci2023nature}, Eq.~\eqref{flatAband} becomes a valid wavefunction right at the magic angle where ${\bm \psi}_{K}(0) = {\bm \psi}_{K'}(0)$ and the pole in $f_{\bf k}$ at $z=0$ is precisely canceled.
The periodic boundary conditions for $ {\bm \psi}_{\bk}$ in momentum space are exactly the same as for the color functions $F^{z}_{\bk,j}$  ($j=1,2$) and the resulting Berry phase integrates to $C=2$ over the moiré Brillouin zone (MBZ) $(\v b_1,\v b_2)$. There are correspondingly~\cite{wang2023origin} two zeroth $k_1$, $k_2$ in the MBZ for ${\bm \psi}_{\bk}$ with the property that their sum is $k_1+k_2 = z$. The two color LLL functions $F^{z}_{\bk,j}$ form a basis for the wavefunctions with two vortices in the MBZ~\cite{wang2023origin}, thereby justifying the decomposition Eq.~\eqref{color_decomposition}. By identification, the space-dependent spinors are related by 
\begin{equation}\label{S-matrix}
    \begin{pmatrix}
        \bm\chi_1 (\br)  e^{-K_1(\v r)} \\ \bm\chi_2 (\br)  e^{-K_2(\v r)}
    \end{pmatrix} = {\cal S} (\v r) \begin{pmatrix}
        {\bm \psi}_{K}(\br) \\ {\bm \psi}_{- K}(\br)
    \end{pmatrix}
\end{equation}
where the coefficients of ${\cal S} (\v r)$ are given in the SM~\cite{Suppmat} (the superscript $z$ is dropped). Each color spinor, $\v \chi_1$ or $\v \chi_2$,  comprises three layers. They transform under SU(3) yielding $8$ components for the layer skyrmions $\bm n_j = \bm \chi^\dagger_j \bm \lambda \bm \chi_j$. The resulting Pontryagin density $Q_1 (\bm r)$ for the first color skyrmion is displayed in Fig.~\ref{fig:figure2}a) and demonstrates a topological winding of $-1$ across the magnetic unit cell $(2 \v a_1,\v a_2)$.  The second color spinor exhibits the same skyrmion texture simply translated by $\v a_1$, $\bm n_2 (\bm r) = \bm n_1 (\bm r + \v a_1)$,  as a consequence of Eq.~\eqref{real-space-BC}. Fig.~\ref{fig:figure2}b) shows the trace of the real-space Berry curvature, periodic over the moiré unit cell spanned by $\v a_1$ and $\v a_2$. 
Finally, we also compute and show in Fig.~\ref{fig:figure2}c) the Wilson loop eigenvalues for the non-Abelian  connection $\bm{\mathcal A} (\bm r)$, defined Eq.~\eqref{berry-connection}. 
The two eigenvalues are exchanged across the moiré unit cell signalling the color entanglement~\cite{wang2023origin}. Their sum winds by $-1$, consistently with the magnetic phase screening.
We emphasize that the magnetic unit cell chosen here is arbitrary.
A rotation in color space defines new layer skyrmions with a different real space profile and periodicities. However, the Wilson loop eigenvalues are independent of this gauge fixing and the total winding is always $-1$ to screen the fictitious magnetic fields in the ideal band wavefunction. We also find a similar skyrmion texture in monolayer-bilayer graphene~\cite{Suppmat}.

\paragraph*{Finite magnetic field and hidden wavefunctions ---} 
Complementing recent studies~\cite{Sheffer_2021,datta2024helical,makov2024flat}, we readily find that our skyrmion decomposition extends to finite (homogeneous) magnetic fields. Eq.~\eqref{color_decomposition} continues to hold and the color LLL functions $F^z_{\v k,j}$  evolve in that case into LLL with a total effective magnetic field 
\begin{equation}\label{sum-magnetic}
B_{\rm eff} = B_0/C + B 
\end{equation}
which is the sum of the fictitious and external magnetic fields. $B_0$ is the homogeneous magnetic field with one flux quantum per unit cell and $B$ the external magnetic field. Eq.~\eqref{sum-magnetic} already predicts the disappearance of the ideal band for a negative flux $\phi = - \phi_0/C$, with $\phi_0 = h/e$ the flux quantum, in agreement with the Streda formula.  In addition, ideal Chern bands are associated with hidden wavefunctions~\cite{Milekhin_2021}: for $C$ colors in a $L$ layer space, there exists $L-C$ (hidden) spinor wavefunctions $\bm\chi_j^{h,z} (\br)$ which span the subspace orthogonal to the color spinors $\bm\chi_j^z(\br)$. They construct an ideal band
\begin{equation}\label{color-decomposition-hidden}
    \bm \psi^h_{\bk}(\br)= \sum_{j=1}^{L-C} \bm\chi_j^{h,z} (\br)\,  F^{h,z}_{\bk,j}(\br) e^{-K^h_j(\v r)}.
\end{equation}
 $F^{h,z}_{\bk,j}(\br)$ are LLL wavefunctions with the effective magnetic field $B_{\rm eff} = -B_0/(L-C)+B$. As a consequence of its holomorphic properties, the wavefunction Eq.~\eqref{color-decomposition-hidden} is defined only for $B_{\rm eff} \ge 0$. 
As the band emerges at the threshold $B = B_0/(L-C)$, the ideal band in the opposite chiral sector disappears~\cite{datta2024helical}. The spinors $\bm\chi_j^{h,z} (\br)$ are not singular but they wind with a {\it positive} Pontryagin index $+1$. The associated real-space boundary conditions are thus obtained from Eq.~\eqref{real-space-BC} with $\v a_j \to - \v a_j$, thereby antiscreening the magnetic field in the LLL functions $F^{h,z}_{\bk,j}(\br)$.
 
\paragraph*{Conclusions ---} Our analysis highlights the layer Skyrme texture of ideal bands to screen the magnetic phase within the Landau level correspondence. 
For higher Chern numbers, the skyrmion decomposition involves a real-space SU(C) Berry connection, associated with a gauge fixing choice of a unit cell. The non-trivial topological winding of layer skyrmions has a direct consequence in single Dirac models: ideal bands are stabilized only when the number of layers exceeds the band Chern number, {\it i.e. $L>C$}. Ideal flat bands emerging at magic angles are spanned by $C$ independent spinors forming a subspace within the $L$-dimensional layer space~\cite{guerci2023nature}.
Interestingly, the properties of higher Chern flat bands $C>1$~\cite{Eslam_highC_idealband,Jie_hierarchyflatband,Yuncheng2023,guerci2023chern,devakul2023magicangle,popov2023magic,popov2023butterfly,Foo2024,kwan2024strongcoupling} in multilayer materials~\cite{xia2023helical,Polshyn_2021,xiaobolu_2024} provide opportunities for predicting new topological phases and excitations beyond quantum Hall analogs~\cite{dong2022,Wu_2015,wang2023origin,wilhelm2023,zhou2024correlatedinsulatorschargedensity}.
Our analysis is further motivated by the recent experimental observation of layer skyrmions in WeS$_2$~\cite{zhang2024directobservationlayerskyrmions}.

\paragraph*{Acknowledgments ---} We are grateful to Nicolas Regnault and Nishchhal Verma for insightful discussions, and to Anushree Datta and Mark Goerbig for a collaboration on a related work. We acknowledge support by the French National Research Agency (project TWISTGRAPH, ANR-21-CE47-0018). The Flatiron Institute is a division of the Simons Foundation.

\bibliography{ref.bib}


\onecolumngrid
\newpage
\makeatletter 

\begin{center}
\textbf{\large Supplementary materials for: `` \@title ''} \\[10pt]
Daniele Guerci$^1$, Jie Wang$^2$, Christophe Mora$^3$   \\
\textit{$^1$ Center for Computational Quantum Physics, Flatiron Institute, 162 5th Avenue, NY 10010, USA}\\
\textit{$^2$Department of Physics, Temple University, Philadelphia, Pennsylvania, 19122, USA}\\
\textit{$^3$ Universit\'e Paris Cit\'e, CNRS,  Laboratoire  Mat\'eriaux  et  Ph\'enom\`enes  Quantiques, 75013  Paris,  France}
\end{center}
\vspace{20pt}

\setcounter{figure}{0}
\setcounter{section}{0}
\setcounter{equation}{0}
\renewcommand{\thetable}{S\arabic{table}}
\renewcommand{\theequation}{S\arabic{equation}}
\renewcommand{\thefigure}{S\arabic{figure}} 

\renewcommand{\thefigure}{S\@arabic\c@figure}
\makeatother



\section{Conventions}
\label{notations}

We choose a system of coordinates where employing complex notation $a_1=e^{-i\pi/2}a$, $a_2=\omega a_1$ and $\omega=e^{2\pi i/3}$. The high symmetry points are in the unit cell are $0$ and $\pm \omega^{j-1}a/\sqrt{3}$. The reciprocal lattice is generated by $b_1=-4\pi \omega/(\sqrt{3}a)$ and $b_2=-\omega^* b_1$. Additionally, we define the high-symmetry points $K=q_1$ and $K'=-q_1$ with $q_1=4\pi e^{i\pi/2}/(3a)$. For an ideal band with Chern number $C$, the fictitious magnetic field of the LLL wavefunction is chosen with the magnetic length $\ell_B$ such that
\begin{equation}
   l^2_B= \frac{C(\v a_1\times \v a_2)}{2 \pi} 
\end{equation}
and a fraction of the flux quantum, $\phi_0/C = h/(e C)$, threads through the unit cell. Throughout this paper, we use the shorthand notation $\v r_1 \cross \v r_2 \equiv (\v r_1 \cross \v r_2) \cdot \v z$ whenever the two vectors $\v r_1$ and $\v r_2$ belong to the 2D plane.


\section{Layer skyrmion: connecting real space Berry curvature with the skyrmion winding}
\label{skyrmion}

\subsection{Real space boundary conditions and Chern number}

We first discuss the case of a single color, relevant for an ideal band with Chern number $C=1$. We employ the notation adopted by Ref.~\cite{DongPRL_2023,JieWang_exactlldescription} which consists of defining the real space Berry connection as $\bm {\mathcal A}=-i\bm\chi^\dagger\nabla_{\br}\bm\chi$ from the normalized spinor wavefunction $\bm  \chi (\bm r)$, with $\bm\chi^\dagger \bm\chi =1$. $\bm \chi (\v r)$ arises from the decomposition of the ideal band given as Eq.~\eqref{ideal_C1_wfc} in the main text. The real-space Chern number $C_R$ is defined from the contour integral
\begin{equation}\label{real_space_C}
    C_{R}=\frac{1}{2\pi}\oint_{\cal C} d\bm l\cdot \bm {\mathcal A} = \frac{1}{2 \pi} \int_{\cal S} d^2 r \, \Omega
\end{equation}
with the oriented path $\cal C$ following the boundaries of the unit cell spanned by $\v a_1$
 and $\v a_2$ and $\cal S$ is the corresponding area. We used Stoke's theorem in Eq.~\eqref{real_space_C} to relate the line integral to the surface integration of the real-space Berry curvature $\Omega = \nabla_{\bm r}\times \bm{\mathcal A}$. We emphasize that the gauge field ${\mathcal A}$ and $\Omega$ characterize the real-space structure of the spinor: they are intrinsically different from the (momentum-space) usual Berry phase of the ideal band. 
 
 The path $\cal C$ consists of four straight lines connected two by two by the vectors $\v a_1$, $\v a_2$ and traversed in opposite directions. The boundary conditions on the gauge field $\bm {\mathcal A}$ are deduced from those of the spinor wavefunctions given in the main text and repeated here for clarity
\begin{equation}\label{Abelian_part_phase}
    \bm\chi(\br+\bm a_j)=e^{i\varphi_{\v a_j}(\br)}\bm\chi(\br), \qquad \varphi_{\v a_j}(\br)=(\br\times\v a_j)/(2l^2_B), \qquad  \bm {\mathcal A} (\bm r + \v a_j) =  \bm {\mathcal A} (\bm r) + \nabla_{\br} \varphi_{\v a_j}(\br).
\end{equation}
The evaluation of the real-space Chern number $C_R$ in Eq.~\eqref{real_space_C} is readily seen to depend only on $\nabla_{\br} \varphi_{\v a_j}(\br)$, with the result
\begin{equation}
     C_{R}= \frac{1}{2\pi} \int^1_0dl [\v a_2\cdot\nabla_{\br}\varphi_{\v a_1}(\br)-\v a_1\cdot\nabla_{\br}\varphi_{\v a_2}(\br)] = \frac{\v a_2\times\v a_1}{2\pi l^2_B}=-1
\end{equation}
We thus find that the screening boundary conditions on the spinor wavefunction, imposed by the LLL factorization, automatically implies a real-space Chern number of $-1$.

The above derivation extends trivially to the case of mutiple colors. The non-Abelian Berry connection is introduced in Sec.~\ref{skyrmionlargerC}, Eq.~\eqref{gauge_field}, and its trace satisfies the real-space boundary conditions
\begin{equation}
    {\rm Tr}   \bm {\mathcal A} (\bm r + \v a_j) =  {\rm Tr}\bm {\mathcal A} (\bm r) +  C \, \frac{\v a_j \times \bm z}{2l^2_B},
\end{equation}
and the real-space Chern number evaluated over the unit cell $(\v a_1,\v a_2)$ is given by
\begin{equation}
     C_{R}=\frac{1}{2\pi}\oint_{\cal C} d\bm l\cdot {\rm Tr} \bm {\mathcal A} = \frac{C (\v a_2\times\v a_1)}{2\pi l^2_B}=-1.
\end{equation}

\subsection{Single-component wavefunction}

We analyze the case where $\chi (\v r)$ has a single component (or layer) and show by contradiction that the LLL decomposition imposes that  $\chi (\v r)$ has to vanish somewhere in the moiré unit cell. Let us rewrite the LLL decomposition as
\begin{equation}
\psi_{\bk}(\br)=\chi(\br)\,e^{-K(\v r)} \, \Phi_{\bk}(\br)
\end{equation}
where $ \Phi_{\bk}(\br)$ is a LLL wavefunction with a homogeneous magnetic field of one flux quantum per unit cell. The fictitious magnetic field inhomogeneity is separated and included in the  real (Kähler) potential $K (\br)$ in such a way that $\chi (\bm r)$ is normalized  $|\chi (\bm r)|^2 =1$ at each position in the unit cell. The wavefunction $\chi (\bm r)$ must be non-vanishing before its normalization by the Kähler potential. In addition, $K (\bm r)$ is periodic over the moiré lattice with  $K (\bm  r + \v a_j) =  K (\bm r)$. The compatibility  between the Bloch and magnetic periodities of $\Phi_{\bk}$ and $\psi_{\bk}$ implies that $\chi (\v r) \equiv e^{i \theta_0 (\v r)}$ transforms under lattice translations as Eq.~\eqref{Abelian_part_phase} with the result
\begin{equation}\label{BC-phase-SC}
    \theta_0 (\v r + \v a_j) = \theta_0 (\v r) + \frac{\br\times\v a_j}{2 l^2_B}.
\end{equation}
The winding of the phase $\theta_0 (\v r)$ along the boundary of the unit cell is $-2 \pi$, as calculated from Eq.~\eqref{BC-phase-SC}, which is consistent with the single valuedness of $\chi (\v r)$. However, this vortex configuration for the phase leads to a contradiction since the path can be contracted to a single point, and $\chi (\v r)$ is of modulus one everywhere on the plane.

In summary, we have proven that the envelope spinor function $\bm \chi (\v r)$, assuming the absolute value does not vanish anywhere in the plane, must consist of at least two components.

\subsection{Layer spinor Skyrme texture and Pontryagin index}

We consider the single color spinor wavefunction $\bm \chi (\v r)$ satisfying the screening boundary condition of Eq.~\eqref{Abelian_part_phase}. The Berry curvature tensor $\Omega_{jk} = -2 {\rm Im} T_{jk}$
is related to the antisymmetric imaginary part of the quantum geometric tensor
\begin{equation}
    T_{jk} = \langle \partial_j \bm \chi | \left( 1 - |  \bm \chi \rangle \langle \bm \chi |\right) | \partial_k \bm \chi \rangle
\end{equation}
where $j$ or $k$ is either $x$ or $y$, and the Berry curvature introduced in Eq.~\eqref{real_space_C} is $\Omega \equiv \Omega_{xy}$. Introducing the projector operator $P (\bm r) = | \bm \chi (\bm r) \rangle \langle \bm \chi (\bm r) |$, the quantum geometric tensor takes the form
\begin{equation}\label{geometric_tensor}
     T_{jk} (\bm r) = {\rm Tr} \left[ \partial_j P  (\bm r) \, \{ 1- P (\bm r) \}  \, \partial_k P (\bm r) \right].
\end{equation}
If the spinor $\chi$ has two components, then its projector can be decomposed into the Pauli matrices of SU(2),
\begin{equation}\label{pauli-decomposition}
    P (\bm r) = \frac{1 +  \bm n (\bm r) \cdot {\bm \sigma}}{2}
\end{equation}
where the normalized vector $\bm n (\br) = \langle \bm \chi (\bm r) | \bm \sigma | \bm \chi (\bm r) \rangle$ is a vector on the Bloch sphere of radius one. Injecting the Pauli matrices decomposition Eq.~\eqref{pauli-decomposition} into the expression of the quantum geometric tensor, one can relate, after some algebra, the real-space Berry curvature
\begin{equation}\label{relation-PI-BC}
    \Omega (\bm r) = 2 \pi Q (\bm r)
\end{equation}
to the Pontryagin density
\begin{equation}\label{pontryagin-density}
    Q (\v r) = \frac{1}{4 \pi}\, \bm n (\bm r) \cdot \Big(\partial_x\bm n (\bm r) \times\partial_y\bm n (\bm r) \Big),
\end{equation}
which describes the rotation of the Bloch vector $\bm n (\bm r)$ as the position $\bm r$ is varied. As this vector is periodic over the moiré lattice, its covering of the Bloch sphere across a moiré unit cell is quantized by the topological Pontryagin index,
\begin{equation}\label{integrated-pontryagin}
    \mathcal W_{\chi} = \int_{{\cal S}\equiv UC} d^2 r \, Q(\v r) = \frac{1}{4 \pi} \int_{ UC} d^2\br \, \, \bm n\cdot(\partial_x\bm n\times\partial_y\bm n) 
\end{equation}
which must be an integer. Using the relation between the Pontryagin index and the real-space Berry curvature, Eq.~\eqref{relation-PI-BC}, and inserting Eq.~\eqref{pontryagin-density} into Eq.~\eqref{real_space_C}, one finally arrives at the result that $\mathcal W_{\chi}$ and $C_R$ are identical. The screening $\mathcal W_{\chi} = C_R = -1$ is thus associated with a skyrmion texture for the Bloch vector $\bm n$ imposed by the Landau level decomposition.

The above arguments can be generalized to a color spinor with $L$ components transforming under SU(L). The decomposition Eq.~\eqref{pauli-decomposition} of the projector~\cite{PhysRevB.104.085114}
\begin{equation}
     P (\v r) = \frac{1}{N} +  \frac{\bm n (\bm  r) \cdot {\bm \lambda}}{2}
\end{equation}
now involves the $L^2-1$ generators of SU(L), and the generalized Bloch vector has also $L^2-1$ components obtained  from  $\bm n_\alpha (\br) = \langle \bm \chi (\bm r) | \lambda_\alpha | \bm \chi (\br) \rangle$. The relation between the real-space Berry curvature and the Pontryagin index Eq.~\eqref{relation-PI-BC} is recovered~\cite{PhysRevResearch.4.023120,PhysRevB.104.085114} as well as Eq.~\eqref{integrated-pontryagin} where the triple product
\begin{equation}
    \bm n\cdot(\partial_x\bm n\times\partial_y\bm n) = f_{abc} n_a(\br) \partial_x n_b(\br) \partial_y n_c(\br) 
\end{equation}
is generalized to SU(L) using the structure factors $f_{abc}$.

\section{Color Landau levels}

We build a basis of lowest Landau levels, with Bloch (pseudo-)periodicity, adapted to describe ideal bands with Chern number $C$. We consider a unit cell with an area such that a single flux quantum threads precisely $C$ unit cells. We thus define the extended magnetic cell 
  $(C \v a_1,\v a_2)$ and the magnetic translations satisfy
\begin{equation}\label{commutationC}
    T_{\v a_1} T_{\v a_2} =  e^{i \phi} \, T_{\v a_2} T_{\v a_1} \qquad \qquad \phi = \frac{e B \v a_1 \cross \v a_2 }{\hbar} = \frac{\v a_1 \cross \v a_2 }{\ell_B^2}  = \frac{2 \pi}{C}
\end{equation}
along with the commutation of $T_{C \v a_1}$ and $T_{\v a_2}$. We have defined 
\begin{equation}
    T_{\v a_1} =  e^{-i \frac{\v b_2}{C} \cdot \v R} \qquad  T_{\v a_2} =  e^{i \frac{\v b_1}{C} \cdot \v R}      
\end{equation}
with the guiding center position
\begin{equation}
     \v R = \v r - \frac{\v z \cross \v \Pi}{e B} \qquad \qquad  \v \Pi = -i \hbar \v \nabla - e \v A
\end{equation}
the reciprocal basis vectors $\v b_1$ and $\v b_2$ and the vector potential $\v A$.
The Bloch magnetic states  $| \v k \rangle = e^{i \v k \cdot \v R} \, | 0 \rangle$ form a basis of periodic Landau levels. They diagonalize the magnetic translation operators $T_{C \v a_1}$ and $T_{\v a_2}$. We define the first color component
\begin{equation}
    F^z_{k,1}(\br) = \langle \bm r | \v k\rangle = U^z_{k,1}(\br) e^{i \bk \cdot \br} ,
\end{equation}
which satisfies the magnetic periodicity
\begin{equation}\label{magnetic-perioC}
    T_{C \v a_1}  U^z_{k,1} (\br) =  U^z_{k,1} (\br), \qquad \qquad T_{\v a_2}  U^z_{k,1} (\br) =   U^z_{k,1} (\br),
\end{equation}
The subsequent color wavefunction are obtained by simply applying the magnetic translation operator $T_{\v a_1}$,
\begin{equation}
    U^z_{k,n}(\br) = T_{\v a_1} U^z_{k,n-1}(\br),
\end{equation}
corresponding to $C$ different colors, due to 
the cyclic condition  $T_{C \v a_1} U^z_{k,1} (\br)= (T_{\v a_1})^C U^z_{k,1} (\br) = U^z_{k,1} (\br)$. The action of the magnetic translation $T_{\v a_2}$ on this color basis is readily obtained from the commutation property Eq.~\eqref{commutationC},
\begin{equation}
    T_{\v a_2} U^z_{k,n}(\br) = e^{i \frac{2 \pi (n-1)}{C}} U^z_{k,n}(\br)
\end{equation}
Building a vector with color components, 
    \begin{equation}
    \bm U^z_k(\br) = \begin{pmatrix}
            U^z_{k,1}(\br) \\ \vdots \\
            U^z_{k,C}(\br)
        \end{pmatrix} 
\end{equation}
The representation of the two magnetic translation operators are $T_{\v a_1} = \sigma$, $T_{\v a_2} = \tau$, with $\sigma$ and $\tau$ given in the main text,
corresponding to the algebra of matrices in the clock or parafermion models,
\begin{equation}
    \sigma^C = \tau^C = 1 \qquad \qquad \sigma \tau = e^{2 i \pi/C} \tau \sigma 
\end{equation}
Remarkably, $\sigma$ and $\tau$ play a symmetric role. The eigenvalues of $\sigma$, $\tau$ and $\sigma^\dagger$ are the same. We define the $x$ axis by a rotation in the color basis which diagonalizes $\sigma$. After this rotation, the roles of $\tau$ and $\sigma$ are simply exchanged and
\begin{equation}
    T_{\v a_1} \bm U^x_k(\br) = \tau \bm U^x_k(\br) \qquad \qquad T_{\v a_2} \bm U^x_k(\br) = \sigma \bm U^x_k(\br)
\end{equation}
generalizing the $\mu_x \leftrightarrow \mu_z$ swap in the half flux case, see Table~\ref{tab:color_basis}.
The magnetic translation along the third symmetric direction $\v a_3$ is obtained from the product
\begin{equation}
    T_{\v a_1} T_{\v a_2} T_{\v a_3} = e^{i \phi/2} = e^{i \pi/C}
\end{equation}
imposed by the flux piercing the oriented triangle formed by $\v a_1$, $\v a_2$ and $\v a_3$. In momentum space, the boundary conditions take a symmetric form
\begin{equation}
     \bm F^z_{\v k+ \v b_1/C} (\br) =  e^{i \phi_{\v k,\v b_1}} \v \, \tau \, \bm F^z_{\v k} (\v r), \qquad  \bm F^z_{\v k+ \v b_2/C} (\br) = e^{i \phi_{\v k,\v b_2} } \v \, \sigma \,\bm F^z_{\v k} (\v r), \qquad \phi_{\v k,\v b_j} = \frac{\ell_B^2}{2 C} \, \v b_j \cross \v k 
\end{equation}
and the rotation to the $x$ axis exchanges $\sigma$ and $\tau$. Applying the momentum transport $C$ times leads to diagonal boundary conditions along both directions,
\begin{equation}
     \bm F^z_{\v k+ \v b_j} (\br) =  e^{i C \phi_{\v k,\v b_1}}  \, \bm F^z_{\v k} (\v r), \qquad \qquad  \bm F^z_{\v k+ \v b_2} (\br) = e^{i C \phi_{\v k,\v b_2} }  \bm F^z_{\v k} (\v r).
\end{equation}
They are sufficient to determine the Chern number for a color LLL over the Brillouin zone $(\v b_1, \v b_2)$
\begin{equation}
       \frac{1}{2 \pi} \left( - \phi_{\v k_0 + \v b_1,\v b_2} + \phi_{\v k_0,\v b_2} - \phi_{\v k_0,\v b_1} + \phi_{\v k_0+ \v b_2,\v b_1} \right) = \frac{\ell_B^2 \v b_1 \cross \v b_2}{2 \pi} = C
\end{equation}
thereby forming a convenient basis to represent Chern $C$ ideal bands. Note that these LLL have Chern $C=1$ when calculated over the $C$ times smaller magnetic Brillouin zone $(\v b_1/C,\v b_2)$. The real space boundary conditions are given in the main text.

\section{Layer skyrmion for $C>1$}
\label{skyrmionlargerC}

\subsection{Non-Abelian Berry connection}

Employing the basis $\bm U^a_{\bk}$, any ideal Chern band with $C>1$ is expressed as~\cite{wang2023origin,dong2022}: 
\begin{equation}
  \bm  \psi_{\bk}(\br)=\sum_{j=1}^C\bm \alpha^a_j(\br) F^a_{\bk,j}(\br),
\end{equation}
where $\bm\alpha^a_{j}$ are spinors in the $L$ dimensional layer space describing the electronic distribution in the heterostructure, we observe that the spinors are not orthogonal $\bm\alpha_n^\dagger\cdot\bm \alpha_{m}\neq 0$ and $\bm\alpha_n^\dagger\cdot\bm \alpha_{n}=N_n\neq 1$. From the magnetic translation properties of $\bm U^a_{\bk}(\br)$ we find: 
\begin{equation}
\begin{split}\label{bc_layer_skyrmions}
    \bm \alpha^z_j(\br+\v a_1)&=e^{i\frac{\br\times\v a_1}{2l^2_B}}U_{\v a_1}\sum_{n=1}^C\bm \alpha^z_{n}(\br)\sigma^\dagger_{nj},\\
    \bm \alpha^z_j(\br+\v a_2)&=e^{i\frac{\br\times\v a_2}{2l^2_B}}U_{\v a_2}\sum_{n=1}^C\bm \alpha^z_{n}(\br)\tau^\dagger_{nj},
\end{split}
\end{equation}
where $U_{\bm a_j}$ is a matrix of phases in the layer degree of freedom which depends on the $\br$ boundary conditions of the moir\'e Hamiltonian, and the boundary condition for $\v a_3$ can be directly obtained from Eq.~\eqref{bc_layer_skyrmions}. Furthermore, the real space boundary conditions for the other directions $\bm \alpha^a$ with $a=x,y$ is obtained by diagonalizing the magnetic translation along $T_{\v a_1}$ and $T_{\v a_3}$, respectively. Notice that we can introduce the normalised spinor $\bm \chi^z_c(\br)=\bm \alpha^z_c(\br)/|N_c(\br)|$ where $|N_c(\br)|=\sqrt{\bm \alpha^{z\dagger}_{c}(\br)\cdot\bm \alpha^z_c(\br)}$ which satisfies the same magnetic translation properties in Eq.~\eqref{bc_layer_skyrmions}.
 
From now on, we set $a=z$ knowing that the other choices are related by unitary transformation. To define the non-Abelian real space Berry connection, we introduce the dual basis $\tilde{\bm \chi}_j$: 
\begin{equation}\label{dual_basis}
    X_{nm}={\bm\chi}_n^\dagger\cdot\bm \chi_m,\quad \tilde{\bm\chi}_n^\dagger= \sum_{m}X^{-1}_{nm}{\bm\chi}_m^\dagger,
\end{equation}
where we dropped the upper index $a$. By definition we have: 
\begin{equation}
    \tilde{\bm\chi}_n^\dagger\cdot {\bm\chi}_m=\delta_{nm},\quad .
\end{equation}
Additionally, we find: 
\begin{equation}
    X_{nm}(\br+\v a_1)= \left[\sigma X(\br)\sigma^\dagger \right]_{nm},\quad    X_{nm}(\br+\v a_2)= \left[\tau X(\br)\tau^\dagger \right]_{nm},
\end{equation}
the same boundary conditions also apply to $X^{-1}(\br)$. we conclude that
\begin{equation}\label{applayer_skyrmion}
    \tilde{\bm \chi}^\dagger_c(\br+\v a_1)=e^{-i\frac{\br\times\v a_1}{2l^2_B}}\sum_{n}\sigma_{cn}\tilde{\bm \chi}^\dagger_{n}(\br)U^\dagger_{\v a_1},\quad    \tilde{\bm \chi}^\dagger_c(\br+\v a_2)=e^{-i\frac{\br\times\v a_2}{2l^2_B}}\sum_{n}\tau_{cn}\tilde{\bm \chi}^\dagger_{n}(\br)U^\dagger_{\v a_2}.
\end{equation}
We now introduce the real space Berry connections: 
\begin{equation}
    \bm{\mathcal A}_{nm}(\br)=-i\tilde{\bm\chi}_n^\dagger(\br)\cdot \nabla_{\br}{\bm\chi}_m(\br).
\end{equation}
This geometrical non-Abelian field is associated to the gauge arbitrariness in the choice of basis color functions, $z$, $x$ or any other rotation in color space. This choice also determines the form of the magnetic unit cell.
Under real space translations, the real-space Berry connection transforms as: 
\begin{equation}\label{real_space_translations_BC}
    \bm{\mathcal A}_{nm}(\br+\v a_1) = \delta_{nm}\nabla_{\br}\varphi_{\v a_1}(\br) + [\sigma \bm {\mathcal A}(\br) \sigma^\dagger]_{nm},\quad  \bm {\mathcal A}_{nm}(\br+\v a_2) = \delta_{nm}\nabla_{\br}\varphi_{\v a_2}(\br) + [\tau \bm {\mathcal A}(\br) \tau^\dagger]_{nm}, 
\end{equation}
where the phase $\varphi_{\v a_{1/2}}(\br)$ has been introduced in Eq.~\eqref{Abelian_part_phase}. Eq.~\eqref{real_space_translations_BC} also implies that the non-Abelian real space Berry connection can be written as: 
\begin{equation}\label{gauge_field}
  \bm{\mathcal  A}(\br)=\frac{\br\times\bm z}{2l^2_B}\lambda_0+\bm{\mathcal A}_0 (\v r)\lambda_0+\sum^{C^2-1}_{j=1} \bm{\mathcal A}_{j}(\br)\lambda_j,
\end{equation}
where $\lambda_j$ are the generators of SU(C) and $\lambda_0$ the identity. $\bm{\mathcal A}_0 (\v r)$ is periodic over the moiré lattice contributing with a spatially modulated field $\nabla_{\br}\times\bm{\mathcal A}_0$ with zero average. We observe that the first two contributions constitute the conventional Abelian part, while the third term arises from the SU(C) non-Abelian structure associated with the internal color degree of freedom.

\subsection{Proof of $C \le L-1$}


For ideal bands in single Dirac models, the number of layers $L$ must exceed the Chern number $C$, $L>C$. This result can be rigorously proven under two key assumptions:
\begin{enumerate}
    \item the layer spinors $\bm{\chi}_n (\br)$ are normalized at each $\br$. This requires that their unnormalized counterparts  $\bm  \alpha_n (\br)$ must not vanish anywhere in the moiré unit cell.
    \item The subspace generated by the spinors $\bm{\chi}_n (\br)$ must have a rank of $C$ at each $\br$. In other words, the spinors must be linearly independent.
\end{enumerate}
They are satisfied for single Dirac models but not always for double Dirac models, see Sec.~\ref{app:QBC} and Refs.~\cite{Eugenio_2023,Wan_prl_2023}. We now proceed with the proof by contradiction and assume that the number of layers equals the Chern number $L=C$.
We employ the ket notation
$\ket{\bm\chi_n}$ for the $L=C$ layer spinor $\bm \chi_n$ with $n=1,\cdots,C$, and define the projector: 
\begin{equation}
    P=\sum_{n=1}^C \ket{\bm\chi_n}\bra{\tilde{\bm \chi}_n},
\end{equation}
satisfying $P^2=P$. 
The assumption 2. implies that the set  $\{\ket{\bm\chi_{1}},\dots,\ket{\bm\chi_{C}}\}$ is linearly independent and thus form a basis for any wavefunction $\ket{\psi}$, expressed as $\ket{\psi}=\sum_{n=1}^{C}\lambda_n \ket{\bm\chi_{n}}$. Applying $P$ gives $P\ket{\psi}=\ket{\psi}$ or $P=\mathbf{1}$, the projector is the identity. We use this result in the expression of the Berry curvature
\begin{equation}\begin{split}
    \Omega(\br)&=\Tr[\nabla_{\br}\times \bm{\mathcal A}(\br)]=-i\sum_n[\braket{\partial_x\tilde{\bm\chi}_n}{\partial_y{\bm\chi}_n}-\braket{\partial_y\tilde{\bm\chi}_n}{\partial_x{\bm\chi}_n}]=-i\sum_n[\braket{\partial_x\tilde{\bm\chi}_n}{\mathbf 1|\partial_y{\bm\chi}_n}-\braket{\partial_y\tilde{\bm\chi}_n}{\partial_x{\bm\chi}_n}]\\
    &=-i\sum_n[\braket{\partial_x\tilde{\bm\chi}_n}{ P|\partial_y{\bm\chi}_n}-\braket{\partial_y\tilde{\bm\chi}_n}{\partial_x{\bm\chi}_n}]=-i\sum_n\left[\sum_m\braket{\partial_x\tilde{\bm\chi}_n}{{\bm \chi}_m}\braket{\tilde{\bm \chi}_m}{\partial_y{\bm\chi}_n}-\braket{\partial_y\tilde{\bm\chi}_n}{\partial_x{\bm\chi}_n}\right]\\
    &=-i\sum_n\left[\sum_m\braket{\partial_y\tilde{\bm \chi}_m}{{\bm\chi}_n}\braket{\tilde{\bm\chi}_n}{\partial_x{\bm \chi}_m}-\braket{\partial_y\tilde{\bm\chi}_n}{\partial_x{\bm\chi}_n}\right]=0,
\end{split}\end{equation}
where, in the last expression, we used the identity $\braket{\tilde{\bm\chi}_n}{{\bm \chi}_m}=\delta_{nm}\implies\braket{\tilde{\bm\chi}_n}{\partial_a{\bm \chi}_m}=-\braket{\partial_a\tilde{\bm\chi}_n}{{\bm \chi}_m}$. 
Alternatively, we arrive at the same conclusion by readily using the definition Eq.~\eqref{geometric_tensor} for the quantum geometric tensor which we find to be identically vanishing when $P=1$. A vanishing Berry curvature throughout the moiré unit cell implies that the real-space Chern number $C_R$ is zero which is fundamentally  incompatible with the screening of the magnetic phase detailed in Sec.~\ref{skyrmion} and in the main text.
Therefore, we have proven by contradiction that $L=C$ is not possible and thus $L-1 \ge C$.


\section{Model Hamiltonian: examples of higher Chern bands}
\label{model_higher_chern}

In this section we illustrate graphene base moir\'e heterostructure realizing ideal $C>1$ bands. Our aim is to expand the ideal wavefunction in the color basis and characterize the properties of the layer skyrmion. 

\subsection{Helical Trilayer Graphene}
\label{hTG}

Focusing on the ABA stacking the Hamiltonian of hTG~\cite{Yuncheng2023} reads: 
\begin{equation}
    H_{\rm ABA}(\br) = \begin{pmatrix}
    v_F \bk\cdot{\bm \sigma}   & T_{\omega^*}(\br) & 0\\
    h.c. & v_F \bk\cdot{\bm \sigma} & T_\omega(\br) \\
    0 & h.c. & v_F \bk\cdot{\bm \sigma}
    \end{pmatrix}.
\end{equation}
The resulting matrices are: 
\begin{equation}
\begin{split}
    &T_{\omega^*}(\br)\equiv T(\br,{\bm\phi})=\alpha\sum^3_{j=1}\left(\omega^*\right)^{j-1}T_je^{-i \bq_j\cdot\br},\\
    &T_{\omega}(\br)\equiv T(\br,-{\bm\phi})=\alpha\sum^3_{j=1}\omega^{j-1}T_je^{-i \bq_j\cdot\br},
    \end{split}
\end{equation}
where ${\bm\phi}=[0,2\pi/3,-2\pi/3]$, $\omega=e^{2\pi i/3}$ and $T_j$ are the matrices
\begin{equation}
\begin{split}
    T_1=\kappa 1+\begin{pmatrix}
        0 & 1 \\
        1 & 0 
    \end{pmatrix},\,T_2=\kappa 1+\begin{pmatrix}
        0 & \omega^* \\
        \omega & 0 
    \end{pmatrix},\,T_3=\kappa 1+\begin{pmatrix}
        0 & \omega \\
        \omega^* & 0 
    \end{pmatrix},
\end{split}
\end{equation}
with $\kappa= w_{\rm AA}/w_{\rm AB}$. We project the model in the basis: 
\begin{equation}
\label{bloch_periodic_htg}
\bm \psi_{\bk} = e^{i\bk\cdot\br}\begin{pmatrix}
        u_{\bk 1 } e^{-i\bq_1\cdot\br} \\ 
        u_{\bk 2 }  \\ 
        u_{\bk 3 } e^{i\bq_1\cdot\br}
    \end{pmatrix},
\end{equation}
leading to the real-space boundary conditions:  
\begin{equation}
    \bm\psi_{\bk}(\br +\bm a_{1/2}) = e^{i\bk\cdot\bm a_1}U^{\bm a_{1/2}}\bm\psi_{\bk}(\br),\quad U^{\bm a_{1/2}}=\text{diag}[\omega^*,1,\omega].
\end{equation}

\begin{figure}
    \centering
    \includegraphics[width=\linewidth]{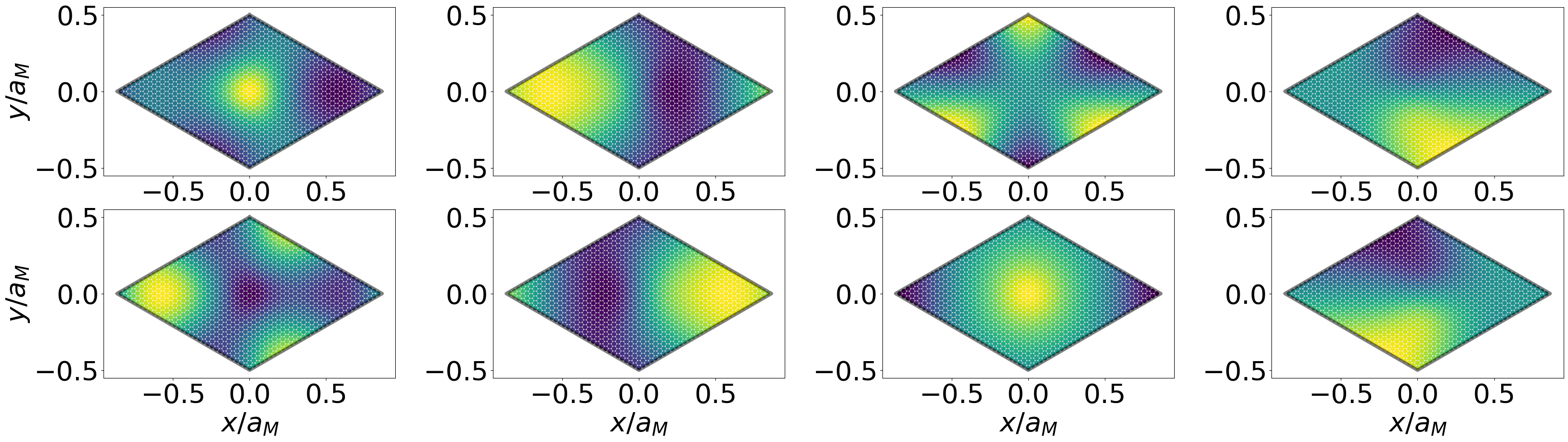}
    \caption{Layer skyrmion for the $C=1$ band of htg BAB stacking configuration. From left to right we show $n_1,\dots,n_8$ with $n_9=1$ since by definition $\lambda_9=1$ and the skyrmion is normalized to 1 for the $C=1$ ideal band.}
    \label{fig:layer_sk}
\end{figure}

\begin{figure}
    \centering
    \includegraphics[width=0.5\linewidth]{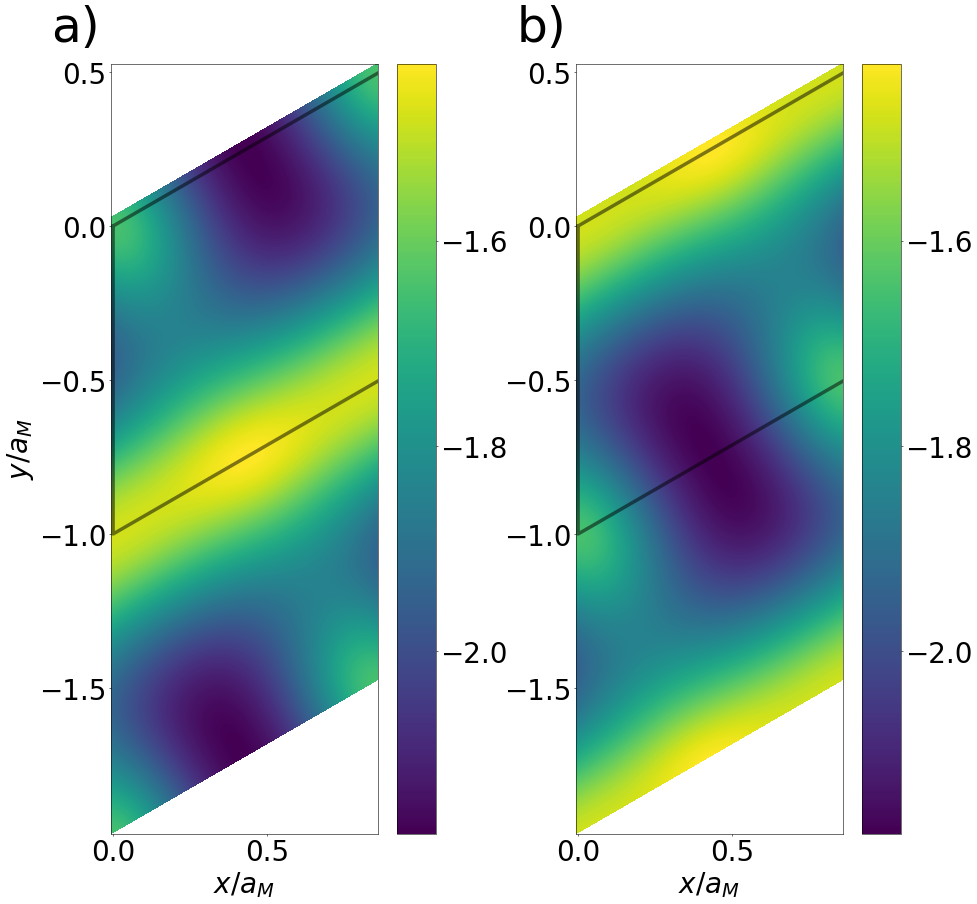}
    \caption{Kahler potential $K_j=-\log|{\bm \alpha}_j|$ for the two different colors composing the $C=2$ ideal wavefunction of htg. Notice that $K_1(\br+\v a_1)=K_2(\br)$. The data are in the enlarged unit cell, while the black line shows the original moir\'e unit cell.}
    \label{fig:kahler_potential}
\end{figure}

\begin{figure}
    \centering
    \includegraphics[width=\linewidth]{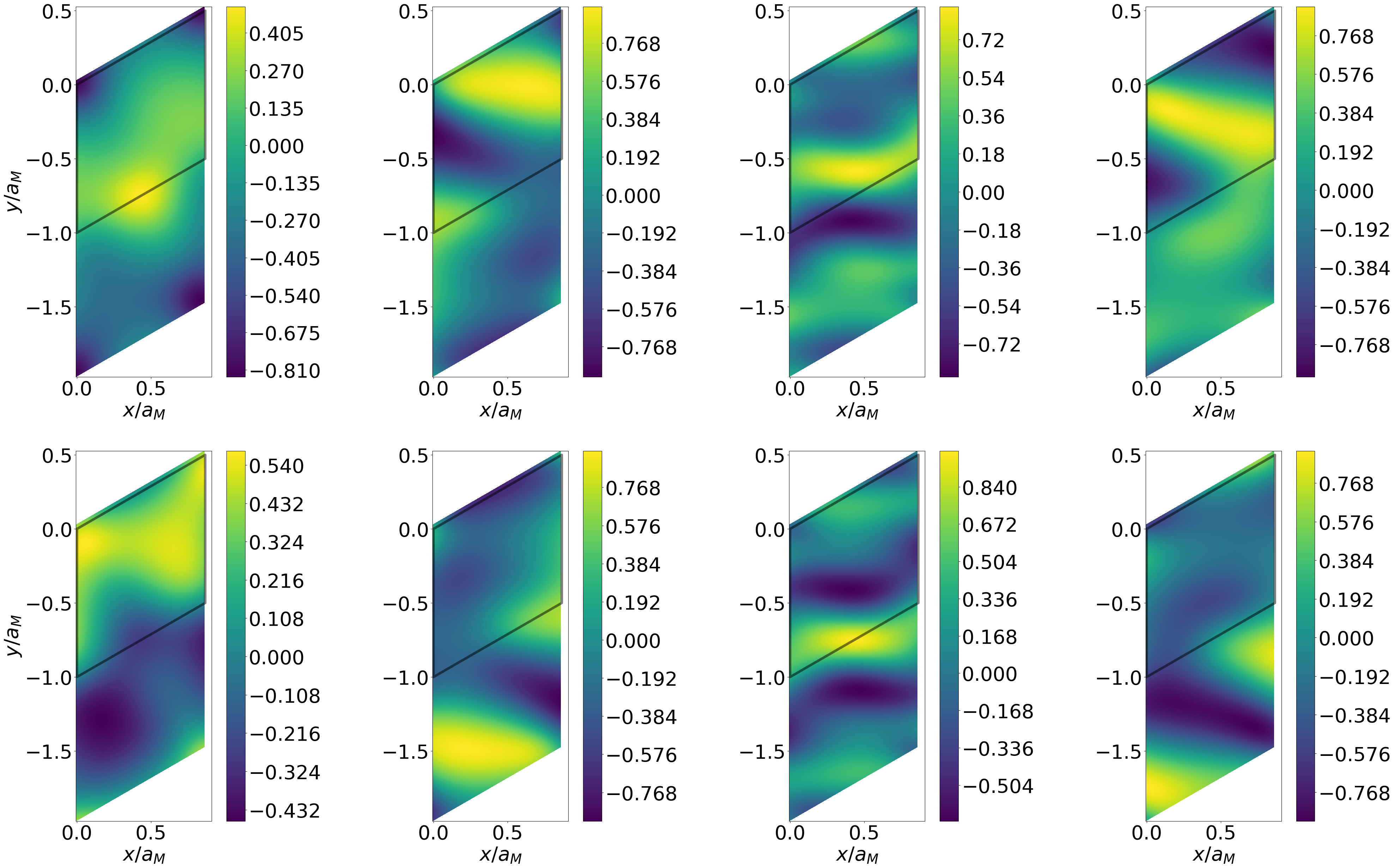}
    \caption{Layer skyrmion of one of the two colors of the $C=2$ band of htg ABA stacking configuration. From left to right we show $n_1,\dots,n_8$ with $n_9=1$ since by definition $\lambda_9=1$ and the skyrmion is normalized to 1.}
    \label{fig:layer_sk_C2}
\end{figure}

In the following we characterize the properties of the ideal flat bands in the chiral limit at the first magic angle, $\theta = 1.687^\circ$ corresponding to $\alpha=0.377$. 
The BAB Hamiltonian is obtained employing $C_{2z}$. We show the components of the layer skyrmion for the $C=1$ band in Fig.~\ref{fig:layer_sk}. We emphasize that the $C=1$ wavefunction vanishes at $\br=0$ for $\bk=\Gamma$. Following Ref.~\cite{Grisha_TBG,Grisha_TBG2,Wang_2021}, the ideal wavefunction is obtained multiplying $\bm\psi_{\Gamma}(\br)$ by the meromorphic function in Eq.~\eqref{holomorphic}:
\begin{equation}\label{C1}
   \bm\psi_{\bk}(\br) = f_{\bk}(z)\bm\psi_\Gamma(\br). 
\end{equation}
The layer skyrmion is then obtained factoring out the lowest Landau level wavefunction $\bm \psi_{\bk}(\br) = \Phi_{\bk}(\br) \bm{\chi}(\br)$. 
Furthermore, we provide an explicit recipe to compute the coefficients of the matrix  ${\cal S}$ in Eq.~\eqref{S-matrix}. The starting point is the ideal band wavefunction expressed as in Ref.~\cite{guerci2023chern}
\begin{equation}
     {\bm \psi}_{\bk}(\br) = R_{\v k,1} (z)  {\bm \psi}_{K}(\br)
     + R_{\v k,2} (z)  {\bm \psi}_{-K}(\br),
\end{equation}
where we employ the notations
\begin{equation}
\begin{split}
    R_{\v k,1} (z) = e^{\frac{\pi}{\rm Im \omega} (k^2 - |k|^2)} a_{k} 
   f_{\bk - \v K} (z), \\
   R_{\v k,2} (z) = e^{\frac{\pi}{\rm Im \omega} (k^2 - |k|^2)} a_{-k} 
   f_{\bk + \v K} (z), 
\end{split}
\end{equation}
and drop the prefactor  ${\cal N}_k$ since it will not play any role in the following discussion. The wavefunction is also given by 
\begin{equation}
    \bm \psi_{\bk}(\br)=  \bm\chi_1^z(\br)\,  F^z_{\bk,1}(\br) e^{-K_1(\v r)} + \bm\chi_2^z(\br)\,  F^z_{\bk,2}(\br) e^{-K_2(\v r)}.
\end{equation}
We can use the boundary condition upon shifting $\v k$ by $\v b_1/2$,
\begin{equation}
      \v F^z_{\v k+ \v b_1/2} (\br) =  e^{i \frac{\ell_B^2}{4} \, \v b_1 \cross \v k} \v \, \mu_z \, F^z_{\v k} (\v r),
\end{equation}
and compute $\bm \psi_{\bk}(\br) \pm \bm \psi_{\bk+ \v b_1/2}(\br) e^{-i \frac{\ell_B^2}{4} \, \v b_1 \cross \v k}$. 
We identify terms and obtain
\begin{equation}\label{S-matrix}
\begin{split}
    \bm\chi_1 (\br)  e^{-K_1(\v r)} = \sum_{j=1}^2 \frac{R_{\v k,j} (z) + \xi_k R_{\v k+\v b_1/2 ,j} (z)}{2 F^z_{\v k,1} (\v r)} {\bm \psi}_{K_j}(\br) \\[2mm]
     \bm\chi_2 (\br)  e^{-K_2(\v r)} = \sum_{j=1}^2 \frac{R_{\v k,j} (z) - \xi_k R_{\v k+\v b_1/2 ,j} (z)}{2 F^z_{\v k,2} (\v r)} {\bm \psi}_{K_j}(\br) 
\end{split}
\end{equation}
with $\xi = e^{-i \frac{\ell_B^2}{4} \, \v b_1 \cross \v k}$. From the latter expression we read out the coefficients of the matrix ${\cal S} (\v r)$:
\begin{equation}
    \begin{split}
       {\cal S}_{1,j} (\v r)   = \frac{R_{\v k,j} (z) + \xi_k R_{\v k+\v b_1/2 ,j} (z)}{2 F^z_{\v k,1} (\v r)},  \\[2mm]
       {\cal S}_{2,j} (\v r)   = \frac{R_{\v k,j} (z) - \xi_k R_{\v k+\v b_1/2 ,j} (z)}{2 F^z_{\v k,2} (\v r)}, 
    \end{split}
\end{equation}
where the coefficients can be verified not to depend on $\v k$ as a result of properties of $\vartheta_1$ functions.
We make a choice of the symmetric gauge for the color wavefunctions with the expressions
\begin{equation}
\begin{split}
F^z_{k,1/2}(\br) = Z_{k,1/2}(z) \,  e^{\frac{\pi}{4 \omega_2} (z^2/a_1^2 - |z/a_1|^2 )} e^{\frac{\pi}{\omega_2} ( k^2/b_2 - |k/b_2|^2 )} e^{i (\bm k + \frac{\v b_1}{4}) \cdot \v a_1 z/a_1} e^{- i \pi k/b_2}
\end{split}
\end{equation}
where we have reintroduced explicitly $a_1$ and $b_2$ unit scales of position $z$ and momentum $k$, $\omega_2 = {\rm Im} \omega$ and the functions
\begin{equation}
\begin{split}
   &Z_{k,1}(z) = i \vartheta_1\left[ \frac{1}{2} + \frac{z}{2 a_1} - \frac{k}{b_2}+\frac{\omega}{4}  ,\frac{\omega}{2}\right] , \\
   &Z_{k,2}(z) = \vartheta_1\left[ \frac{z}{2 a_1} - \frac{k}{b_2}+\frac{\omega}{4}  ,\frac{\omega}{2}\right] 
\end{split}
\end{equation}
The Kahler potential associated with the $C=2$ band of ABA htg is shown in Fig.~\ref{fig:kahler_potential}a) and Fig.~\ref{fig:kahler_potential}b) for color 1 and 2, respectively. Fig.~\ref{fig:layer_sk_C2} shows the different components of $\bm n_c=\bm \chi_c\bm \lambda\bm \chi_c$ for $c=1$.

\subsection{Monolayer-Bilayer Graphene}
\label{mdBLG}

The Hamiltonian reads
\begin{equation}
    H(\br) = \begin{pmatrix}
        v_F\bk\cdot\bm\sigma & T_0 & 0  \\ 
        T^\dagger_0 & v_F\bk\cdot\bm\sigma & T(\br)  \\ 
        0 & T^\dagger(\br) & v_F\bk\cdot\bm\sigma \\ 
    \end{pmatrix},
\end{equation}
the model has three free dimensionless parameters: 
\begin{equation}
    \alpha =\frac{w_{\rm AB}}{\hbar v_F k_\theta} , \quad \kappa= \frac{w_{\rm AA}}{w_{\rm AB}}, \quad \beta = \frac{\gamma}{\hbar v_Fk_\theta},
\end{equation}
and 
\begin{equation}
\label{tunneling_zero}
    T_0=\gamma\begin{pmatrix}
        0 & 0 \\ 
        1 & 0 
    \end{pmatrix}.
\end{equation}
Additionally, we have: 
\begin{equation}
    T(\br) = \alpha \sum_{j=1}^3 T_j e^{-i\bq_j\cdot\br}.
\end{equation}
We project the Hamiltonian in the basis 
\begin{equation}
\label{bloch_periodic}
\bm \psi_{\bk} = e^{i\bk\cdot\br}\begin{pmatrix}
        u_{\bk 1 }  \\ 
        u_{\bk 2 }  \\ 
        u_{\bk 3 } e^{i\bq_1\cdot\br}
    \end{pmatrix}
\end{equation}
where $u_{\bk \ell}$ are cell periodic and we have fixed two Dirac cones at $\Gamma$ and one at $K'$. 
The definition Eq.~\eqref{bloch_periodic} gives rise to the space boundary conditions: 
\begin{equation}
    \bm\psi_{\bk}(\br +\bm a_{1/2}) = e^{i\bk\cdot\bm a_1}U^{\bm a_{1/2}}\bm\psi_{\bk}(\br),\quad U^{\bm a_{1/2}}=\text{diag}[1,1,\omega].
\end{equation}
In the chiral limit $\kappa=0$ in the sublattice basis the model takes the off-diagonal form: 
In the chiral limit where 
\begin{equation}
    H = \begin{pmatrix}
        0 & \mathcal D^\dagger \\ 
        \mathcal D  & 0 
    \end{pmatrix}_{AB},
\end{equation}
the zero-mode operator takes the form
\begin{equation}\label{zero-mode-operator}
    \mathcal D = \begin{pmatrix}
        - 2 i \bar{\partial} & \beta & 0 \\
        0  & - 2 i \bar{\partial} & \alpha U_\omega(\br) \\
        0 & \alpha U_\omega (-\br) & - 2 i \bar{\partial} 
    \end{pmatrix}
\end{equation}
$\bar\partial=(\partial_x+i\partial_y)/2$ and $U_{\omega} (\br) = e^{-i \bq_1 \cdot \br} + \omega e^{-i \bq_2 \cdot \br} + \omega^* e^{-i \bq_3 \cdot \br} $. The last two lines form a twisted bilayer graphene coupled by $\beta$ to a single semi-Dirac operator $-2 i \bar{\partial}$. The $C_{3 z}$ symmetry implies that the first two layer-components of the A-sublattice solution $(\psi_1,\psi_2,\psi_3)$ 
\begin{equation}
    \bm\psi_{1,\pm \bq_1} (\br_0) = 0 \qquad  \bm\psi_{2,\pm \bq_1} (\br_0) = 0
\end{equation}
vanish for arbitrary $\alpha$ at $\br_0 = (a_1 - a_2)/3$, while the third component $\psi_{3,\pm \bq_1} (\br_0) \ne 0$ is not vanishing, both at momenta $\bk = \bq_1$ and $-\bq_1$. The energies of the two central bands are vanishing at $\bk = -\bq_1$ (the original position of the bottom Dirac point) but not at $\bk = \bq_1$. As a result, the band is not flat for general $\alpha$. The first band flattening occurs at the twisted bilayer graphene magic angle $\alpha = 0.586$, where the two central bands have zero energy at $\bk = \bq_1$. As a result, writing
\begin{equation}
    \psi_{\bq_1} (\br_0) = \begin{pmatrix}
        0 \\ 0 \\ \Psi_1 
    \end{pmatrix} \qquad \qquad    \psi_{-\bq_1} (\br_0) = \begin{pmatrix}
        0 \\ 0 \\ \Psi_{-1} 
    \end{pmatrix} 
\end{equation}
one realizes that the two vectors $\psi_{\bq_1} (\br)$ and $\psi_{-\bq_1} (\br)$ become colinear at $\br = \br_0$ but with different amplitudes $\Psi_{1/-1}$. The wavefunction in Eq.~\eqref{flatAband} is a zero mode solution with two differences. The first one concern the center of mass position of the zero modes which is now located at $z_0=a_1\sum_{j=1}^{L}\kappa_L/b_2$ where two zero modes at $\kappa_{1,2}=\gamma$ and one at $\kappa_\kappa'$ and we reintroduced explicitly $a_1$ and $b_2$. As a result the meromorphic function constituting the building block of the zero mode wavefunction reads 
\begin{equation}\label{holomorphic-mod}
   {\bar f}_{\bk} (z) = e^{i \bm k\cdot\v a_1 z/a_1} \frac{\vartheta_1[ (z-z_0)/a_1 - k/b_2,\omega]}{\vartheta_1[ (z-z_0)/a_1,\omega]},
\end{equation}
and
\begin{equation}\label{flatBband}
    \psi_{\bk}(\br) ={\cal N}_k e^{\frac{\pi}{\rm Im \omega} (k^2/b^2_2 - |k/b_2|^2)} \left[ a_k \bar f_{k-q_1} (z) \bm \psi_{\bm q_1} (\br) + b_k \bar f_{k+q_1} (z) \bm \psi_{-\bm q_1} (\br) \right]
\end{equation}
where $a_k$ and $b_k$ are chosen to eliminate the pole at $z=z_0$, or
\begin{equation}
    a_k = \vartheta_1[(q_1+k)/b_2,\omega] \, \Psi_{-1} e^{i \bq_1 \cdot {\bf a}_1 z_0/a_1} \qquad \qquad b_k = \vartheta_1[(q_1-k)/b_2,\omega] \, \Psi_{1} e^{-i \bq_1 \cdot {\bf a}_1 z_0/a_1}.
\end{equation}
The fact that the band flattening occurs precisely at the same value as for TBG, $\alpha = 0.586$, can be understood by making contact with the TBG zero-mode solution. The last two lines of Eq.~\eqref{zero-mode-operator} are identical to finding the zero modes of TBG~\cite{Eslam_highC_idealband}. Hence, the zero-energy solution can also be written~\cite{Eslam_highC_idealband} under the form - up to a prefactor,
\begin{equation}\label{relation-TBG}
   2 i \bar{\partial} \psi_{1,\bk} = t_1 \lambda_{\bk} \psi^{\rm TBG}_{1,\bk}  \qquad  \psi_{2,\bk} = \lambda_{\bk} \psi^{\rm TBG}_{1,\bk}  \qquad  \psi_{3,\bk} = \lambda_{\bk} \psi^{\rm TBG}_{2,\bk}  
 \end{equation}
 for the A sublattice, with  $\lambda_{\bk} = \vartheta_1 [-k/b_2,\omega]$. 
 The wavefunctions of the flat band of twisted bilayer graphene are given by
 \begin{equation}\label{flat-band-TBG}
     \begin{pmatrix}
         \psi^{\rm TBG}_{1,\bk} (\br) \\ \psi^{\rm TBG}_{2,\bk}(\br)
     \end{pmatrix} = \bar f_{\bk} (z) \begin{pmatrix}
         \psi^{\rm TBG}_{1,K}(\br) \\ \psi^{\rm TBG}_{2,K}(\br)
     \end{pmatrix}
 \end{equation}
with the function $\bar f_{\bk} $ given above as Eq.~\eqref{holomorphic-mod}. It is not obvious at first reading why Eq.~\eqref{flat-band-TBG} combined with the relations Eq.~\eqref{relation-TBG} are compatible with our analytical expression Eq.~\eqref{flatBband}. It can however be shown explicitly with the help of the following identity on the $\vartheta_1$ functions
\begin{equation}
\begin{split}
      & \vartheta_1\left( \frac{z-z_0}{a_1} - \frac{k-q_1}{b_2},\omega \right) \vartheta_1\left( \frac{z-z_0}{a_1} - \frac{q_1}{b_2},\omega \right) \vartheta_1\left(   \frac{k+q_1}{b_2},\omega \right) 
      +  \vartheta_1\left( \frac{z-z_0}{a_1} - \frac{k+q_1}{b_2},\omega \right) \\[3mm] & \times \vartheta_1\left( \frac{z-z_0}{a_1} + \frac{q_1}{b_2},\omega \right) \vartheta_1\left(   \frac{k-q_1}{b_2},\omega \right) = \vartheta_1\left( \frac{z-z_0}{a_1} - \frac{k}{b_2},\omega \right) \vartheta_1\left( \frac{z-z_0}{a_1},\omega \right) \frac{\vartheta_1\left(  \frac{2 q_1}{b_2},\omega \right)}{\vartheta_1\left( \frac{q_1}{b_2},\omega \right)}
\end{split}
\end{equation}
Finally, the layer skyrmions composing the color entangled flat band can be readily obtained from the $\mathcal S(\br)$ matrix which can be obtained employing Eq.~\eqref{S-matrix} upon a redefinition of $R_{\bk,j}(z)$ consistent with Eq.~\eqref{flatBband}.

\section{non-Abelian Berry phase and Wilson loop}
\label{Wloop}

In this section we provide the expression of the real space topological properties of the layer skyrmions $\bm \chi_c(\br)$ for a color entangled wavefunction. Employing the dual basis defined in Eq.~\eqref{dual_basis}, in the limit of small $d\br$: 
\begin{equation}
     \tilde{\bm\chi}_n^\dagger(\br)\cdot {\bm\chi}_m(\br+d\br)\simeq\delta_{nm}+id\br\cdot\bm{\mathcal A}_{nm}(\br)\simeq \left[e^{i\int_{\br}^{\br+d\br}d\bm l\cdot\bm{\mathcal A}}\right]_{nm}.
\end{equation}
Considering a closed loop $\mathcal C$ with corners [1,\dots,4] we have: 
\begin{equation}
    \Xi_{nm} = \left[\prod_{j=1}^{4}\tilde{\bm \chi}^\dagger(\br_j)\cdot\bm \chi(\br_{j+1})\right]_{nm}=\left[e^{i\oint_{\mathcal C} d\bm l\cdot\bm {\mathcal A}}\right]_{nm},
\end{equation}
where we emphasize that $O_{nm}(j)=\tilde{\bm \chi}^\dagger_n(\br_j)\cdot\bm \chi_{m}(\br_{j+1})$ is a matrix in the color degree of freedom. The total Berry phase accumulated in the closed loop reads: 
\begin{equation}
    \varphi = \sum_{n=1}^C\frac{\log \lambda_n}{2\pi i},
\end{equation}
where $\lambda_n$ are the eigenvalues of $\Xi_{nm}$. We conclude observing that the phase $\varphi$ is directly related to the Abelian part of the real space Berry curvature $\Omega=\Tr\left[\nabla_{\br}\times\bm{\mathcal A}\right]\approx 2\pi\varphi/(|\v a_1\times \v a_2|/N)$ with $N$ number of points sampling the unit cell $(\v a_1,\v a_2)$. An additional quantity we employ to characterize the topological properties of the layer skyrmion is the Wilson loop: 
\begin{equation}
    \begin{split}
    W_{nm}(x_2)=&\sum_{\{j_{l}\}}\tilde{\bm \chi}^\dagger_n(x_2\v a_2+\v a_1)\cdot\bm \chi_{j_{N-1}}(x_2\v a_2+(1-1/N)\v a_1)\dots\tilde{\bm \chi}^\dagger_{j_{1}}(x_2\v a_2+\v a_1/N)\cdot\bm\chi_{m}(x_2\v a_2).
    \end{split}
\end{equation}
Employing the boundary conditions in Eq.~\eqref{applayer_skyrmion} we have: 
\begin{equation}
    \begin{split}
    W_{nm}(x_2)=e^{ix_2\pi/C}\sum_{\{j_{l}\}}\sum_j\sigma_{nj}\tilde{\bm \chi}^\dagger_j(x_2\v a_2)\cdot\bm \chi_{j_{N-1}}(x_2\v a_2+(1-1/N)\v a_1)\dots\tilde{\bm \chi}^\dagger_{j_{1}}(x_2\v a_2+\v a_1/N)\cdot\bm\chi_{m}(x_2\v a_2),
    \end{split}
\end{equation}
where the matrix of phases has been absorbed in the definition of $\bm \chi$. Similarly, integrating along $\v a_2$ we have: 
\begin{equation}
    \begin{split}
    W_{nm}(x_1)=e^{-ix_1\pi/C}\sum_{\{j_{l}\}}\sum_j\tau_{nj}\tilde{\bm \chi}^\dagger_j(x_1\v a_1)\cdot\bm \chi_{j_{N-1}}(x_1\v a_1+(1-1/N)\v a_2)\dots\tilde{\bm \chi}^\dagger_{j_{1}}(x_1\v a_1+\v a_2/N)\cdot\bm\chi_{m}(x_1\v a_1).
    \end{split}
\end{equation}
From the latter definitions we find that the Wilson loop satisfies the boundary conditions $\hat W(x_1+1)=e^{-i\pi/2}[\tau \hat W(x_1)\tau^\dagger]$ and $\hat W(x_2+1)=e^{i\pi/2}[\sigma \hat W(x_2)\sigma^\dagger]$.

\section{Exact $A$-sublattice zero mode at an high-symmetry momentum in the chiral limit}
\label{chiral_A_sub}

At an high-symmetry $\bk$ point the eigenvalue equation of hTG and twisted mono-double bilayer graphene for the $A$ sublattice zero mode in the chiral limit can be solved exactly. This result also pointed out in Ref.~\cite{makov2024flat} originates from a destructive interference in the momentum space reciprocal lattice that we discuss explicitly in this section. Focusing on hTTG the $C=2$ $A$-sublattice zero mode at $\Gamma$ in the chiral limit, the eigenvalue problem reads: 
\begin{equation}
    \begin{split}
        &(-\bq_1-\bm Q)\cdot\bm\sigma \bm\varphi_{\bm Q,+}+\alpha\sum^3_{j=1}(\sigma^-+\omega^{j-1}\sigma^+)\bm\varphi_{\bm Q+\bm g_j,0}=\epsilon\bm\varphi_{\bm Q,+},\\
        &-\bm Q\cdot\bm\sigma \bm\varphi_{\bm Q,0}+\alpha\sum^3_{j=1}(\sigma^++\omega^{*j-1}\sigma^-)\left(\bm\varphi_{\bm Q-\bm g_j,+}+\bm\varphi_{\bm Q+\bm g_j,-}\right)=\epsilon\bm\varphi_{\bm Q,0},\\
        &(\bq_1-\bm Q)\cdot\bm\sigma \bm\varphi_{\bm Q,-}+\alpha\sum^3_{j=1}(\sigma^-+\omega^{j-1}\sigma^+)\bm\varphi_{\bm Q+-\bm g_j,0}=\epsilon\bm\varphi_{\bm Q,-},
    \end{split}
\end{equation}
where $\bm\varphi_{\bm Q,\ell}$ with $\ell=+,0,-$ is a two dimensional spinor for the upper, middle and lower layer. We know solve the previous equation for the first shell of recoprocal lattice vectors which includes the amplitudes $\bm \varphi_{a,\pm}$ and $\bm\varphi_0$ corresponding to the central site in the middle layer and the six first nearest neighbors: 
\begin{equation}
        \begin{split}
        &-\bm q_a\cdot\bm\sigma \bm\varphi_{a,+}+\alpha(\sigma^-+\omega^{a-1}\sigma^+)\bm\varphi_{0}=\epsilon\bm\varphi_{a,+},\\
        &\alpha\sum^3_{j=1}(\sigma^++\omega^{*a-1}\sigma^-)\left(\bm\varphi_{a,+}+\bm\varphi_{a,-}\right)=\epsilon\bm\varphi_{0},\\
        &-\bm q_a\cdot\bm\sigma \bm\varphi_{a,-}+\alpha(\sigma^-+\omega^{a-1}\sigma^+)\bm\varphi_{0}=\epsilon\bm\varphi_{a,-}.
    \end{split}
\end{equation}
We now look at the $\epsilon=0$ and $A$-sublattice polarized eigenstate. The latter satisfy the simple Schr\"odinger equation: 
\begin{equation}
    \begin{split}
&        \alpha\psi_0+q_a\psi_{a,-}=0,\\
&        \alpha\psi_0-q_a\psi_{a,+}=0, \\
&           \alpha\sum^3_{a=1}\omega^{*a-1}(\psi_{a,+}+\psi_{a,-})=0.
    \end{split}
\end{equation}
One can readily realize that the zero mode solution to this equation reads: 
\begin{equation}
\label{exact_gamma}
    \bm \psi =\frac{e^{i\phi}}{\sqrt{1+6\alpha^2}} \begin{pmatrix}
        &-i\sqrt{3}\bm\varphi_{\omega^*}\\
        &1\\
        &i\sqrt{3}\bm\varphi_{\omega^*}
    \end{pmatrix},
\end{equation}
where $\bm\varphi_{\omega^*}=[1,\omega^*,\omega]^T/\sqrt{3}$ is an eigenstate of $C_{3z}$ with eigenvalues $\omega^*$ and $e^{i\phi}$ is the global U(1) phase. Remarkably, this is the exact as, due to a destructive interference effects, the zero mode wavefunction has vanishing amplitude on the middle layer second shell of momenta. This can be directly checked plugging Eq.~\eqref{exact_gamma} in: 
\begin{equation}
    \begin{split}
&\alpha(\psi_0+\psi_{5,0}+\psi_{6,0})+q_1\psi_{1,-}=0,\\
&\alpha(\psi_0+\psi_{1,0}+\psi_{2,0})+q_2\psi_{2,-}=0,\\
&\alpha(\psi_0+\psi_{3,0}+\psi_{4,0})+q_3\psi_{3,-}=0,\\
&\alpha(\psi_0+\psi_{2,0}+\psi_{3,0})-q_1\psi_{1,+}=0,\\
&\alpha(\psi_0+\psi_{4,0}+\psi_{5,0})-q_2\psi_{2,+}=0,\\
&\alpha(\psi_0+\psi_{1,0}+\psi_{6,0})-q_3\psi_{3,+}=0,\\
&           \alpha\sum^3_{a=1}\omega^{*a-1}(\psi_{a,+}+\psi_{a,-})=0,\\
&\alpha(\omega^*\psi_{3,+}+\omega\psi_{2,-})-g_1\psi_{1,0}=0,\\
&\alpha(\omega^*\psi_{1,+}+\psi_{2,-})-g_2\psi_{2,0}=0,\\
&\alpha(\psi_{3,-}+\omega\psi_{1,+})-g_3\psi_{3,0}=0,\\
&\alpha(\omega^*\psi_{3,-}+\omega\psi_{2,+})-g_4\psi_{4,0}=0,\\
&\alpha(\psi_{2,+}+\omega^*\psi_{1,-})-g_5\psi_{5,0}=0,\\
&\alpha(\omega\psi_{1,-}+\psi_{3,+})-g_6\psi_{6,0}=0,\\
    \end{split}
\end{equation}
where in the latter expression $g_j$ are reciprocal lattice vectors. We conclude that the Eq.~\eqref{exact_gamma} is an exact solution Schr\"odinger equation since additional sites are decoupled $\psi_{1-6,0}=0$. Also Eq.~\eqref{exact_gamma} is confirmed by numerical inspection. To conclude that the mechanism at hand is based on a destructive interference in momentum space as the reciprocal lattice model consists of a dice lattice with complex phases $\exp(\pm2i\pi/3)$~\cite{Vidal_2001}.

\section{Ideal Chern band in a monolayer: quadratic band touching under periodic strain}
\label{app:QBC}

In this section we detail the case of a quadratic band touching under periodic strain discussed in Refs.~\cite{Eugenio_2023,Wan_prl_2023} originating from time-reversal invariant high-symmetry points. Due to time reversal, strain enters as $\partial^2\to\partial^2+A$ leading, in the chiral limit, to the zero mode equation for the A sublattice: 
\begin{equation}
    (4\bar\partial^2+\tilde A^*)\psi_{\bk}=0,
\end{equation}
where $\tilde A=\gamma\sum_{n=1}^{3}\omega^{n-1}\cos(\bm g_n\cdot\br)$ with $\bm g_n$ $C_{3z}$-related reciprocal lattice vectors and $\tilde A(C_{3z}\br)=\omega\tilde A(\br)$. The structure of the zero mode equation, which cannot be reduced to the square of a Dirac operator, gives rise to ideal Chern bands even in the absence of any layer degree of freedom. This phenomenon stems from the higher-order winding characterizing the quadratic band touching and can be understood by expanding the wavefunction at $\bk=\gamma$ around $\bk=0$. Thanks to the $C_{3z}$ and time-reversal symmetry, one readily find that for $\br\to0$
\begin{equation}
    \psi_\gamma(\br)=C_0+C_1 z \bar z+\cdots,
\end{equation}
with $C_n\in\mathbb R$. Expanding the vector potential to the second order in $\br$ we find that the zero mode equation around $\gamma$ reads: 
\begin{equation}
    (4\bar\partial^2+2\pi^2 z^2)\psi_{\gamma}=0\implies C_n=-\frac{\pi^2}{2n(n-1)}C_{n-2},\quad \psi_\gamma=\sum_{n} C_n (z\bar z)^n.
\end{equation}
Tuning the amplitude of the potential $\gamma$ realizes the magic condition $C_0=0$ implying for $\br\to0$: 
\begin{equation}
    \psi_\gamma = C_1 z\bar z+\cdots.
\end{equation}
As a result,  we can construct a zero mode solution for all $\bk$ described by the ideal wavefunction~\cite{Grisha_TBG}: 
\begin{equation}
    \psi_{\bk}(\br)=\Phi^*(\br)\Phi_{\bk}(\br),
\end{equation}
with $\Phi^*(\br)$ anti-Landau level and  $\Phi_{\bk}(\br)$ generalized LLL experiencing an inhomogeneous magnetic field of one flux quantum. Here, $\Phi(\br)$ displays zero at $z=0$ and $\Phi_{\bk}(\br)$ at $z=k a_1/b_2$ in the moir\'e unti cell, respectively.

\end{document}